\documentclass[journal=mamobx,manuscript=letter,email=true,layout=twocolumn]{achemso}

\usepackage[font=footnotesize,labelfont=bf,labelsep=period]{caption}   

\setkeys{acs}{articletitle = true}
\PassOptionsToPackage{english}{babel}
\usepackage[version=3]{mhchem}
\usepackage{abstract}
\usepackage[version=3]{mhchem}      
\usepackage[T1]{fontenc}            
\usepackage[]{graphicx}             
\usepackage{subfig}
\usepackage{epstopdf}
\usepackage{color}
\PassOptionsToPackage{english}{babel}
\usepackage{amssymb,amsfonts,amsmath,cleveref}
\usepackage{newtxmath}
\usepackage{bbold}
\usepackage[normalem]{ulem}
\usepackage{float}
\usepackage{morefloats}
\usepackage{soul}
\usepackage[english]{babel}
\usepackage{multirow}
\usepackage{placeins}
\usepackage{makecell}
\usepackage{xcolor}

\SectionNumbersOn

\author{Alexandros J. Tsamopoulos}
\author{Zhen-Gang Wang}

\affiliation[Caltech]
{Division of Chemistry and Chemical
 Engineering, California Institute of Technology, Pasadena,
 California 91125, United States}
\email{zgw@caltech.edu}

\title[An \textsf{achemso} demo]
  {Ion Conductivity in Salt-Doped Polymers: Combined Effects of Temperature and Salt Concentration  }

\abbreviations{IR,NMR,UV}

\begin{document}


\begin{abstract}

We construct a coarse--grained molecular dynamics model based on poly(ethylene oxide) and lithium bis-(trifluoromethane)sulfonimide salt to examine the combined effects of temperature and salt concentration on the transport properties. 
Salt doping notably slows down the dynamics of polymer chains and reduces ion diffusivity, resulting in a glass transition temperature increase proportional to the salt concentration.
The polymer diffusion is shown to be well represented by a modified Vogel--Fulcher--Tamman (M-VFT) equation that accounts for both the temperature and salt concentration dependence. 
Furthermore, we find that at any temperature, the concentration dependence of the conductivity is well described by the product of its infinite dilution value and a correction factor accounting for the reduced segmental mobility with increasing salt concentration.  
These results highlight the important role of polymer segmental mobility in the salt concentration dependence of ionic conductivity for temperatures near and above the glass transition.


\end{abstract}


Electrolytes in energy storage devices, such as Li-ion batteries, typically consist of organic solvents and binary salts.
Existing commercial energy storage devices use organic electrolytes, which come with safety and performance concerns,\cite{peled_reviewsei_2017} such as thermal runaway and electrolyte decomposition. 
Solvent-free polymer electrolytes (PEs), e.g., poly(ethylene)oxide (PEO), offer promising alternatives with low flammability, good mechanical stability, and suppression of the formation of Li dendrite growth.\cite{goodenough_challenges_2010} 
However, PEs suffer from low ionic conductivity ($\sigma \sim 10^{-4}$ S cm$^{-1}$) and low cation transference number ($t_+ \sim 0.3$).\cite{hallinan_polymer_2013} 
To overcome these performance bottlenecks, a comprehensive understanding of the ion transport mechanisms in these materials is necessary.

Seminal steps in this direction were taken by Borodin and Smith, \cite{borodin_mechanism_2006} who conducted molecular dynamics (MD) simulations with polarizable force fields.
They proposed three ion transport mechanisms, namely: (1) polymer--ion co-diffusion, (2) ion motion along the polymer chain (intra-chain), and (3) inter-segmental ion hopping. 
 Their study indicated that the dynamics of both cations and anions are coupled to the polymer segmental relaxation.  
 Further studies by Maitra and Heuer \cite{maitra_cation_2007, maitra_understanding_2007} provided  analytical expressions for the lithium ion diffusivity $D_{\text{Li}^+}$ in terms of the polymer chain length based on the three characteristic timescales of the proposed ion transport mechanisms.  
Additional MD studies by Molinari et al. \cite{molinari_effect_2018} suggested that polymer segmental relaxation is mainly responsible for the $\text{Li}^+$ diffusion, and that interchain hops are rare; 
however, this conclusion was based on simulations at much higher temperatures than the glass transition temperature of the system.
While these studies focused on the mechanisms that determine $\text{Li}^+$ diffusion, they did not explore the effect of salt concentration on the polymer mobility and on the glass transition temperature, factors that are crucial in determining ion transport. 

Balsara and co-workers \cite{mongcopa_relationship_2018, mongcopa_segmental_2020} conducted quasi-elastic neutron scattering experiments to understand how salt concentration influences the ionic conductivity of PEO doped with lithium bis(trifluoromethane) sulfonimide salt (PEO/LiTFSI). 
They found that the conductivity reaches a maximum with respect to salt concentration, as a result of the competing effects of increased charge carriers and the slowdown of polymer segmental dynamics. This result was corroborated by the computer simulation study of Webb et al. \cite{webb_globally_2018} for PEO doped with LiPF$_6$, where addition of the lithium salt was shown to result in a global slowdown of chain dynamics.


An analogous trend to the concentration dependence of the ionic conductivity was reported in terms of the polymer host polarity. 
Ganesan and coworkers \cite{wheatle_effect_2018, wheatle_effect_2020} performed molecular dynamics simulation to examine how the polar nature of the polymer chain influences ion transport properties. 
It was found that by increasing the dielectric permittivity of the polymer medium, the ionic conductivity increases due to the weakened ionic interactions. 
At higher dielectric constant, however, the enhanced polymer--polymer dipolar interactions impede the segmental relaxation dynamics, leading to a decrease of the ionic conductivity. 
They suggested that there is an intermediate dipole strength for which the ionic conductivity is optimal. 

In this Letter, we report results from coarse-grained molecular dynamics study aimed at exploring the combined effect of salt concentration and temperature on the ion transport in systems such as PEO doped with LiTFSI.
We find that the polymer segmental dynamics are primarily responsible for the non-monotonic behavior of conductivity with salt concentration and propose a modified Vogel--Fulcher--Tamman (M-VFT) equation to describe the temperature and salt concentration dependence of the polymer segmental mobility, that remains valid near the glass transition temperature. 

We follow the simulation framework developed by Hall and co-workers \cite{shen_effects_2020,shen_ion_2020,shen_molecular_2021} to model lithium-salt doped polymers (such as the PEO/LiTFSI system) in which the polymer dynamics is described by the Kremer--Grest model \cite{grest_efficient_1996,kremer_dynamics_1990} and the strong ion--polymer interaction is captured by a $1/r^4$ solvation potential.  
In order to capture glass transition and the effects of temperature on the segmental dynamics, we include an attractive interaction between non-bonded monomer beads. \cite{morita_study_2006,hsu_coarse-grained_2019}
More details on the simulation methods and the mapping from coarse-grained to real units, are provided in the Supporting Information.   

\begin{figure}[h]
\centering
\includegraphics[width=3.25 in]{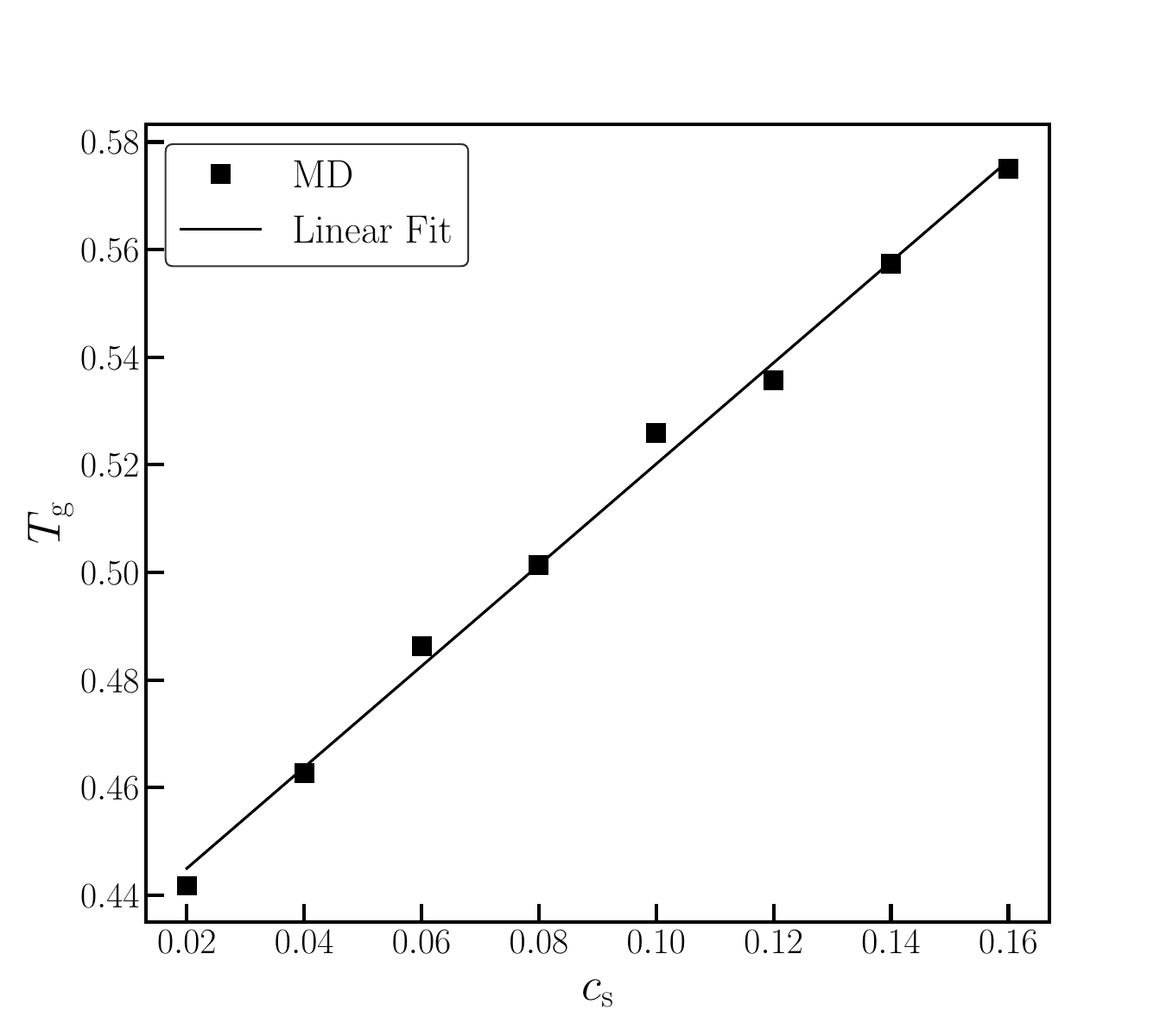}
\caption{Glass transition temperature, $T_{\text g}$, as a function of salt concentration, $c_{\text s}$.}
\label{fig:Tg_cs}
\end{figure}

As the chain dynamics is intimately related to the proximity to glass transition \cite{morita_study_2006} and lithium-salt doping is known to affect the chain dynamics,\cite{webb_globally_2018, mongcopa_relationship_2018} we first examine the dependence of glass transition temperature, $T_{\text{g}}$, on salt concentration. 
A common method for determining $T_{\text{g}}$ is by measuring the temperature dependence of the simulation volume under isobaric conditions. \cite{buchholz_cooling_2002}
At the glass transition, there is a characteristic change in the slope of the volume when plotted as a function of the temperature (see fig S1a).
As shown in \Cref{fig:Tg_cs}, $T_\text{g}$ increases with salt concentration and follows a straight line for the concentration range we studied.
We find that both the magnitude of the $T_{\text {g}}$ shift and the nearly linear dependence on salt concentration are consistent with the experimental findings. \cite{lascaud_phase_1994}.
Considering the simplicity of the coarse-grained model used in our work, such agreement is quite reassuring.  
This behavior can be attributed to the increased ion--polymer complexation leading to restricted motion of the  polymer backbone, \cite{webb_globally_2018} and to the decreased partial molar volume of the polymer (see fig. S1b).
Further, it has been reported in previous studies \cite{nest_crosslinked_1988} that the ion--polymer interactions lead to dynamic cross--linking of the polymer chains, effectively increasing the apparent molecular weight, and thus increasing $T_\text{g}$.\cite{fox_second-order_1950}


The most direct transport properties of the salt-doped polymer system are the self-diffusivities of the ions and center-of-mass self-diffusivity of the polymer, determined from:

\begin{equation}\label{eq:MSD}
D_\alpha = \lim_{t \to \infty} \dfrac{\langle  \Delta \boldsymbol{R}_\alpha(t)^2 \rangle  }{6t}
\end{equation}

\noindent where $ \langle  \Delta \boldsymbol{R}_{\alpha}(t)^2 \rangle$ is the mean--square displacement (MSD) of species $\alpha$ ($\alpha=+$, $-$, and ${\text p}$) during time $t$.  For the polymer, $\boldsymbol{R}_{\text p}(t)$ refers to the center-of-mass position of the chain.
The salt concentration dependence of the diffusion of all species at $T=1.0$ is shown in Figure S2. 
The diffusivity of all species decrease as $c_\text{s}$ increases due to the strong ion--polymer coupling.\cite{borodin_mechanism_2006}

\begin{figure} 
\centering
  \includegraphics[width=3.25 in]{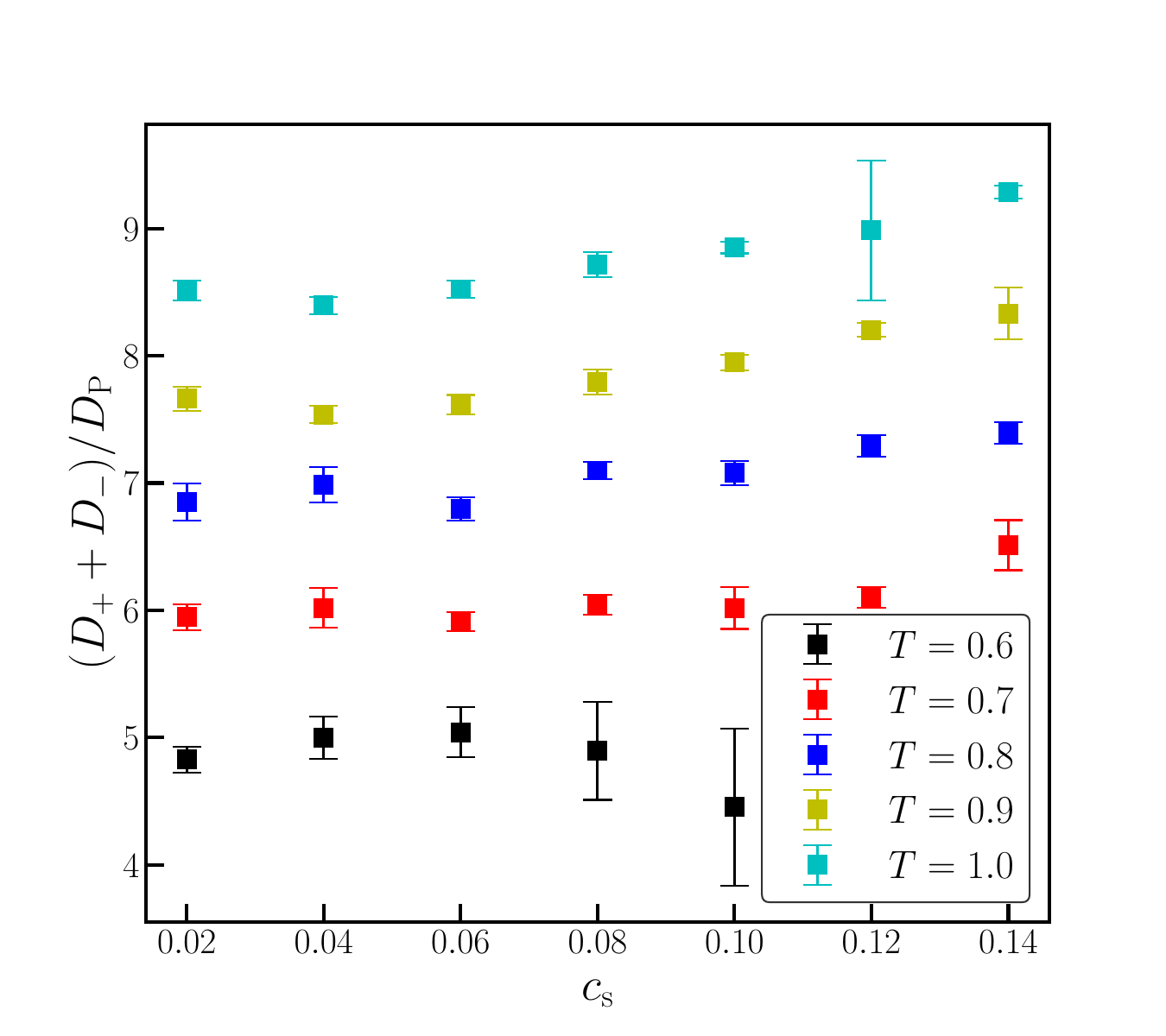}
\caption{Ratio of the total ion diffusion over the polymer diffusion, $(D_++D_-)/D_{\text{P}}$, as a function of salt concentration, $c_{\text{s}}$.}
\label{fig:dratio}
\end{figure}

Since the sum of the anion and cation diffusivities is proportional to the Nernst--Einstein conductivity, we examine its dependence on the salt concentration.
Notably, while all the diffusivities decrease with increasing salt concentration, \Cref{fig:dratio} shows that the ratio $(D_++D_-)/D_{\text{p}} $ is relatively constant for each temperature. 
This result indicates that the ion diffusivities and polymer diffusivity have different temperature effects but similar salt concentration effects. 


For unentangled polymer melts, the polymer diffusivity is inversely proportional to the monomeric friction coefficient $\zeta$. 
Previous studies have employed the Rouse model to examine the dynamics of polymer chains under lithium salt doping, nanoparticle doping, and concurrent salt and nanoparticle doping. \cite{webb_globally_2018,mogurampelly_influence_2016,lin_chain_2021} 
As shown in Figure S4, the chain dynamics is well described by the Rouse modes for all the modes well above the glass transition. 
Near the glass transition, the high-$p$ modes exhibit notable deviation from the expected Rouse dynamics, but the low-$p$ modes still follow the Rouse behavior.  
We thus take the lowest Rouse mode corresponding to the center-of-mass diffusion to define the friction coefficient via $D_{\text{P}} = kT/N \zeta$.
Mongcopa et al.\cite{mongcopa_relationship_2018} have suggested that the friction coefficient follows an exponential dependence on salt concentration.  
Our results confirm this to be the case at high temperatures.  However, at lower temperatures closer to $T_{\text{g}}$, there is strong deviation from the exponential dependence; the friction instead becomes super-exponential (see fig S5). 

Since the diffusion of a pure polymer melt close to $T_{\text{g}}$ is well described by a Vogel--Fulcher--Tamman (VFT) temperature dependence, we propose a modified VFT (M-VFT) formula by
taking into account the observed linear shift in the glass transition temperature with salt concentration and possible increase in the activation energy: \cite{diederichsen_compensation_2017,chintapalli_structure_2016}

\begin{equation}\label{eq:Dcom}
D_{\text{P}} =  B \exp \left[-\dfrac{E^{(0)}+E^{(1)}c}{T-(T_0^{(0)}+a^{(1)}c)} \right] 
\end{equation} 

\noindent where $B$ is a prefactor, $E^{(0)}$ is the activation energy in the pure polymer melt, and 
$T_0^{(0)}$ is the Vogel temperature of the pure polymer melt, \cite{diederichsen_compensation_2017} at which the configurational entropy vanishes \cite{garca-coln_theoretical_1989} (typically taken to be 50K below the glass transition temperature $T_{\text g}$). 
$E^{(1)}$ and $a^{(1)}$ are coefficients for capturing the leading-order dependence on salt concentration in the activation energy and glass transition temperature, respectively.  

\begin{figure}[H]
\centering
\includegraphics[width=3.25 in]{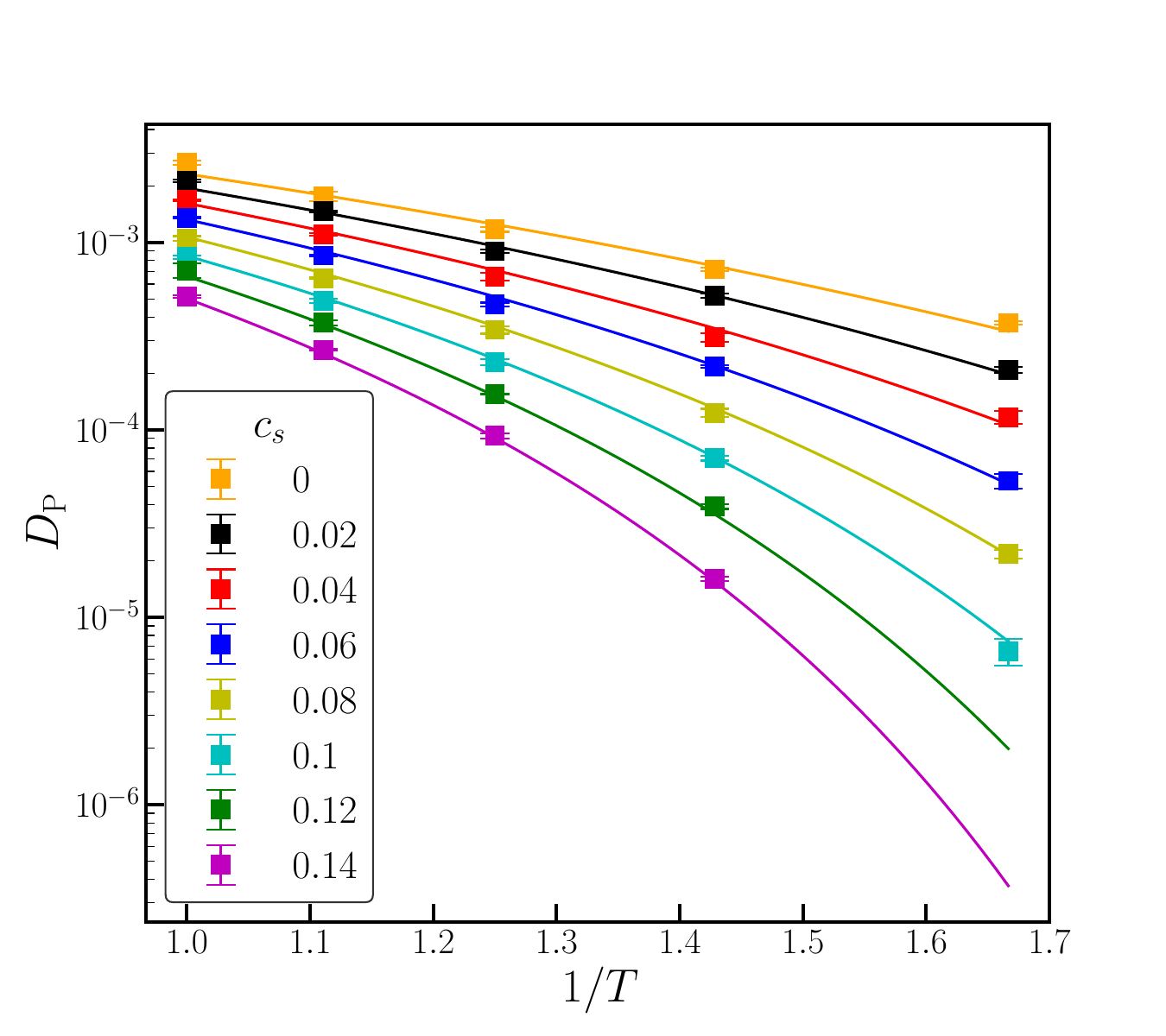}
\caption{Polymer center of mass diffusivity, $D_{\text{P}}$, as a function of 1/T, for various salt concentrations.}
\label{fig:Dcom_cs_T}
\end{figure}

In \Cref{fig:Dcom_cs_T}, we show $D_\text{P}$ in an Arrhenius plot for the 8 salt concentrations (including 0 salt) studied in our work. The symbols are the simulation data.  
$D_{\text{P}}$ is seen to decrease with increasing salt concentration and decreasing temperature. 
The solid lines are the results of fitting. We find that with a single set of fitting parameters, $B =0.01273$, $E^{(0)} =1.295$, $E^{(1)} =4.201$, $T_0^{(0)} =0.245$, $a^{(1)} =1.249$, \cref{eq:Dcom} is able to describe all the simulation data well. 

The ionic conductivity is a complex function of ion--ion correlations and segmental dynamics. 
Typically in an electrolyte solution, with increasing salt concentration, ion--ion correlation in the form of ion pairing and ion clustering increases, leading to reduced ion mobility. \cite{zhang_direct_2015, mcdaniel_ion_2018} 
However, for the PEO--LiTFSI system, the experimental work by Mongcopa et al. \cite{mongcopa_relationship_2018} showed that the reduction of the conductivity is primarily due to the slowing down of the polymer segmental dynamics, implying that ion--ion correlation plays a relatively minor role in governing the conductivity.\cite{son_ion_2020}  
The physical reason for the weak ion--ion correlation in the LiTFSI--PEO system is likely due to the strong Li--PEO interaction which results in the seclusion of the Li$^+$ ions from forming ion pairs and clusters with the bulky TFSI$^-$ ions. \cite{mao_structure_2000}
In our coarse-grained model, the weakness of the ion--ion correlation can be gleaned from the radial distribution function of the polymer beads and anions from the cation (see fig. S3).
To quantify the effects of the ion--ion correlation on the ionic conductivity, we compare its value computed using the exact (Einstein) equation

\begin{equation}\label{eq:sigma_Einstein}
\sigma = \lim_{t \to \infty} \dfrac{e^2}{6tVk_\text{B}T} \sum_{\text{i,j}}^{n} \left\langle \big( z_\text{i} \Delta \boldsymbol{R}_{\text{i}} (t) \big) \cdot  \big( z_{\text{j}} \Delta \boldsymbol{R}_{\text{j}} (t) \big) \right\rangle
\end{equation} 
with the approximate expression from using the ion diffusivities (the Nernst--Einstein equation)

\begin{equation}\label{eq:sigma_Nernst_Einstein}
\sigma^{\text{NE}} = \dfrac{e^2}{Vk_\text{B}T}(N_+z_+^2D_+ + N_-z_-^2D_- )
\end{equation}

\noindent In these expressions, $e$ is the elementary charge, $V$ is the simulation volume, $k_\text{B}$ is the Boltzmann constant, $T$ is the temperature, $z_\text{i}$ is the ion valency, and  $\Delta \boldsymbol{R}_{\text{i}}(t)$ is the displacement of particle i during time $t$. $N_+$ and $N_-$ are respectively the number of cations and anions, and $D_+$ and $D_-$ are the corresponding diffusivity.
\color{black}

\begin{figure}
\centering
\includegraphics[width=3.25 in]{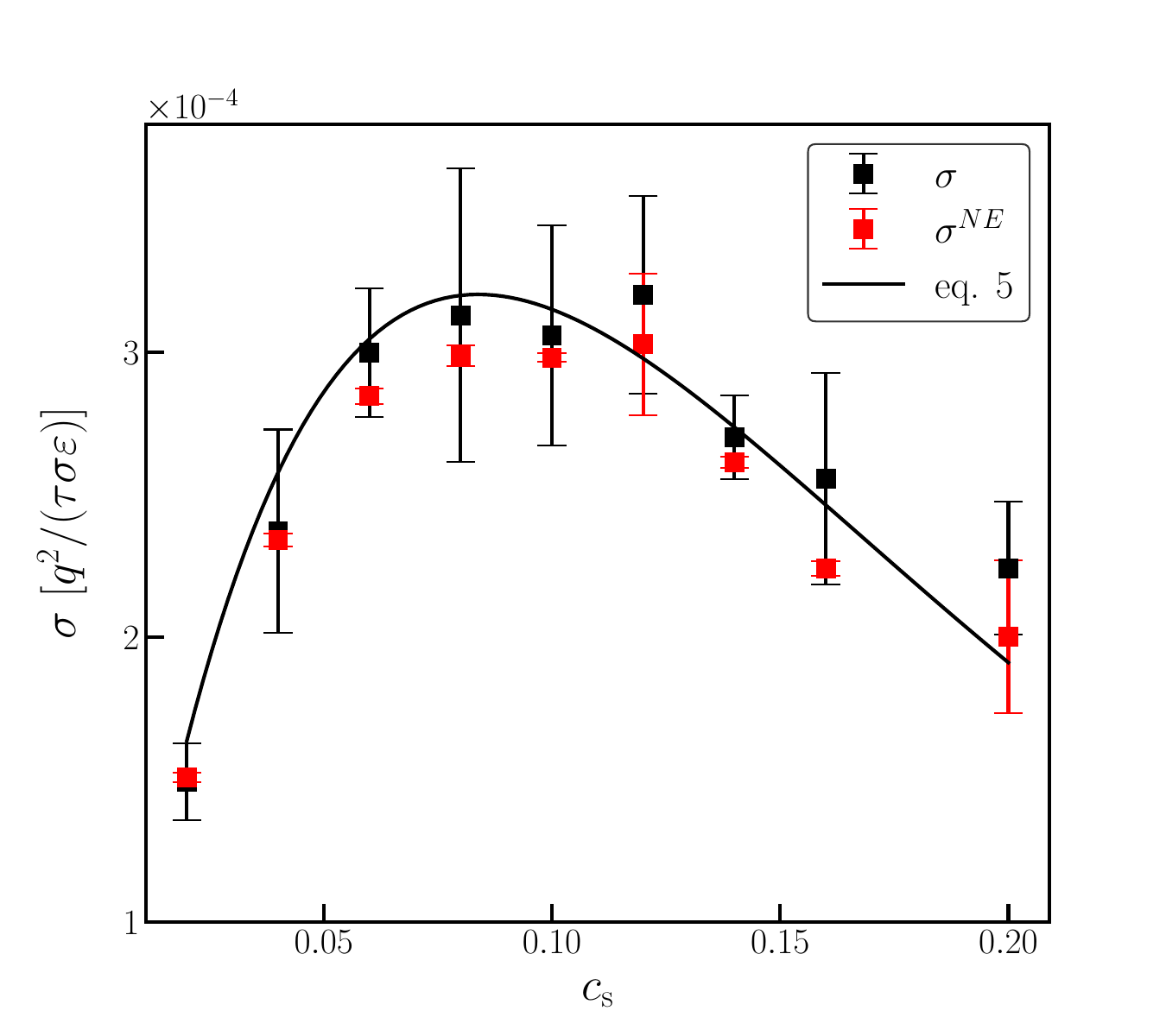}
\caption{Ionic conductivity, $\sigma$, as a function of salt concentration, $c_{\text s}$, at temperature $T=1.0$.  We include the true conductivity, $\sigma$, the Nernst--Einstein conductivity, $\sigma^{\text{NE}}$, and the conductivity based on \cref{eq:Sigma_scale}.
}
\label{fig:cond_cs}
\end{figure}

In \Cref{fig:cond_cs}, we show the ionic conductivity computed using \cref{eq:sigma_Einstein} together with the Nernst--Einstein conductivity calculated using \cref{eq:sigma_Nernst_Einstein} as a function of salt concentration at the high temperature $T=1.0$. 
With increasing salt concentration, the conductivity initially increases, reaches a maximum, and then decreases; this non-monotonic behavior is consistent with both earlier simulations and experiments. \cite{borodin_mechanism_2006,mongcopa_relationship_2018}
Within the error bars of the simulation data, the Nernst--Einstein conductivity is very close to the true conductivity, confirming that ion--ion correlation is unimportant for this system.

Based on the fitted exponential dependence of the friction coefficient at high temperatures, Ref. \citenum{mongcopa_relationship_2018} proposed the following expression for the salt concentration dependence of the conductivity:

\begin{equation}\label{eq:Sigma_scale}
\sigma = \lambda c_\text{s} \text{exp} \left( -\dfrac{c_\text{s}}{c_\text{s}^\text{max}} \right)
\end{equation}
where $c_\text{s}^\text{max}$ is obtained from fitting the friction coefficient.  Our data at $T=1.0$ shown in \Cref{fig:cond_cs} can be reasonably described by this expression. 
Later work by the same group \cite{hoffman_temperature_2021} found that at lower temperatures, \cref{eq:Sigma_scale} could still fit the experimental data; however, the coefficients $\lambda$ and $c_\text{s}^\text{max}$ were treated as purely fitting parameters {\em for each temperature} and no connection was made with the segmental friction.

Since the conductivity is well approximated by the Nernst--Einstein conductivity, and since the ratio of the sum of the cation and anion diffusivities to the polymer diffusivity remains essentially independent of the salt concentration for a given temperature, we hypothesize that using the M-VFT equation (\cref{eq:Dcom}) for the polymer diffusivity would allow a unified description of the conductivity for any temperature and salt concentration.  
Thus, we propose

\begin{equation}\label{eq:sigma_Dcom}
\sigma =  \lambda(T) c_\text{s} \dfrac{D_\text{{P}}(c_\text{s})}{D_\text{{P}}(0)}  
\end{equation} 
where $D_\text{{P}}(c_\text{s})$ is given by \cref{eq:Dcom} and $D_\text{{P}}(0)=D_\text{{P}}(c_\text{s}=0)$. $\lambda(T)$ corresponds to the specific conductivity at infinite dilution at temperature $T$, which can be directly calculated from performing simulations at very low salt concentrations, or treated as a fitting parameter. 
In Figure S8, we show data for $\lambda(T)$ obtained using both methods, with good agreement.  For our numerical fitting using 
\cref{eq:sigma_Dcom}, we adopt the value obtained from fitting. 

\begin{figure}[H]
\centering
\subfloat[]{
\label{fig:cond_cs_T}}
  \includegraphics[width=3.25 in]{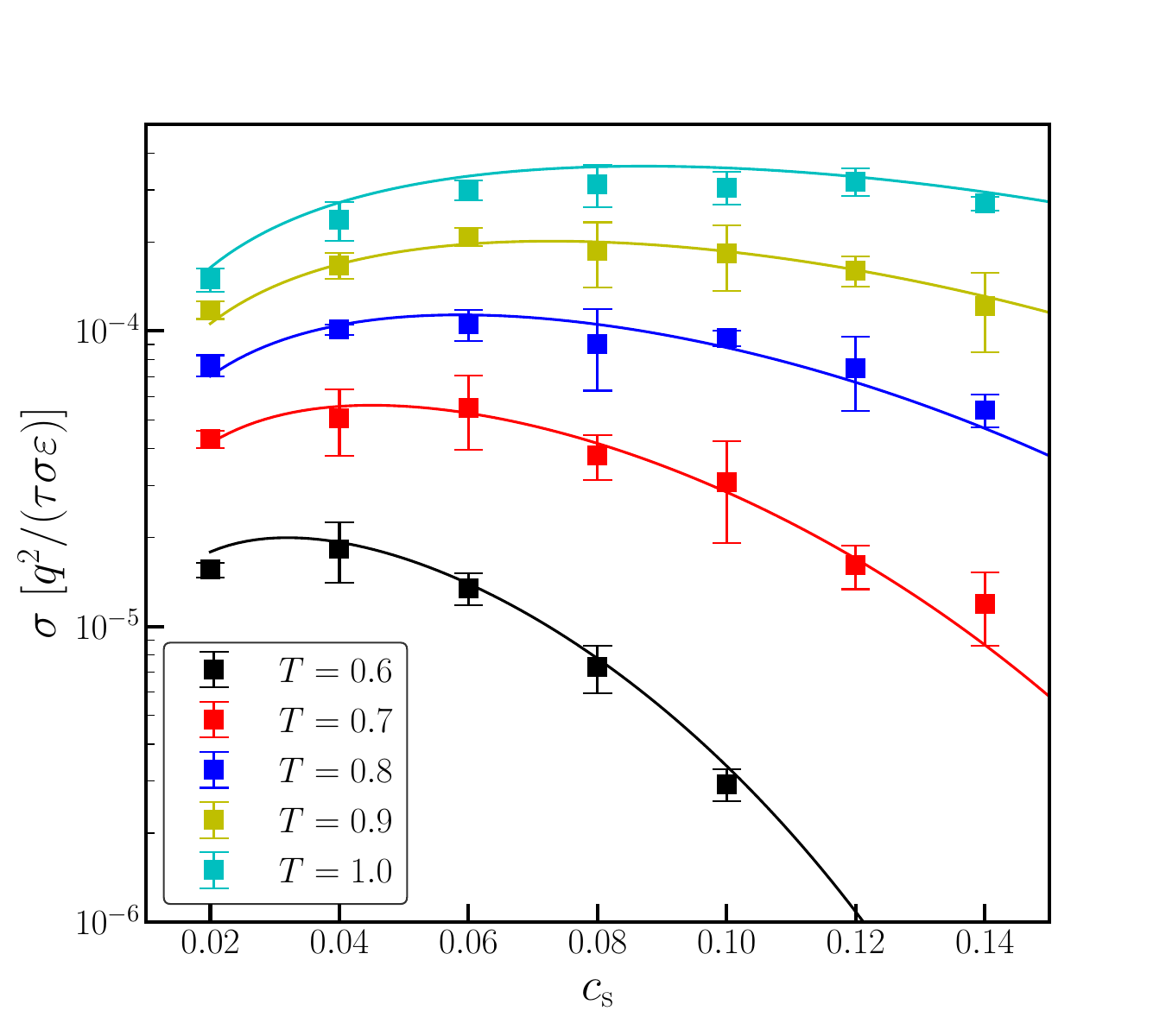}
\subfloat[]{
\label{fig:sigma_D_ratios}}
  \includegraphics[width=3.25 in]{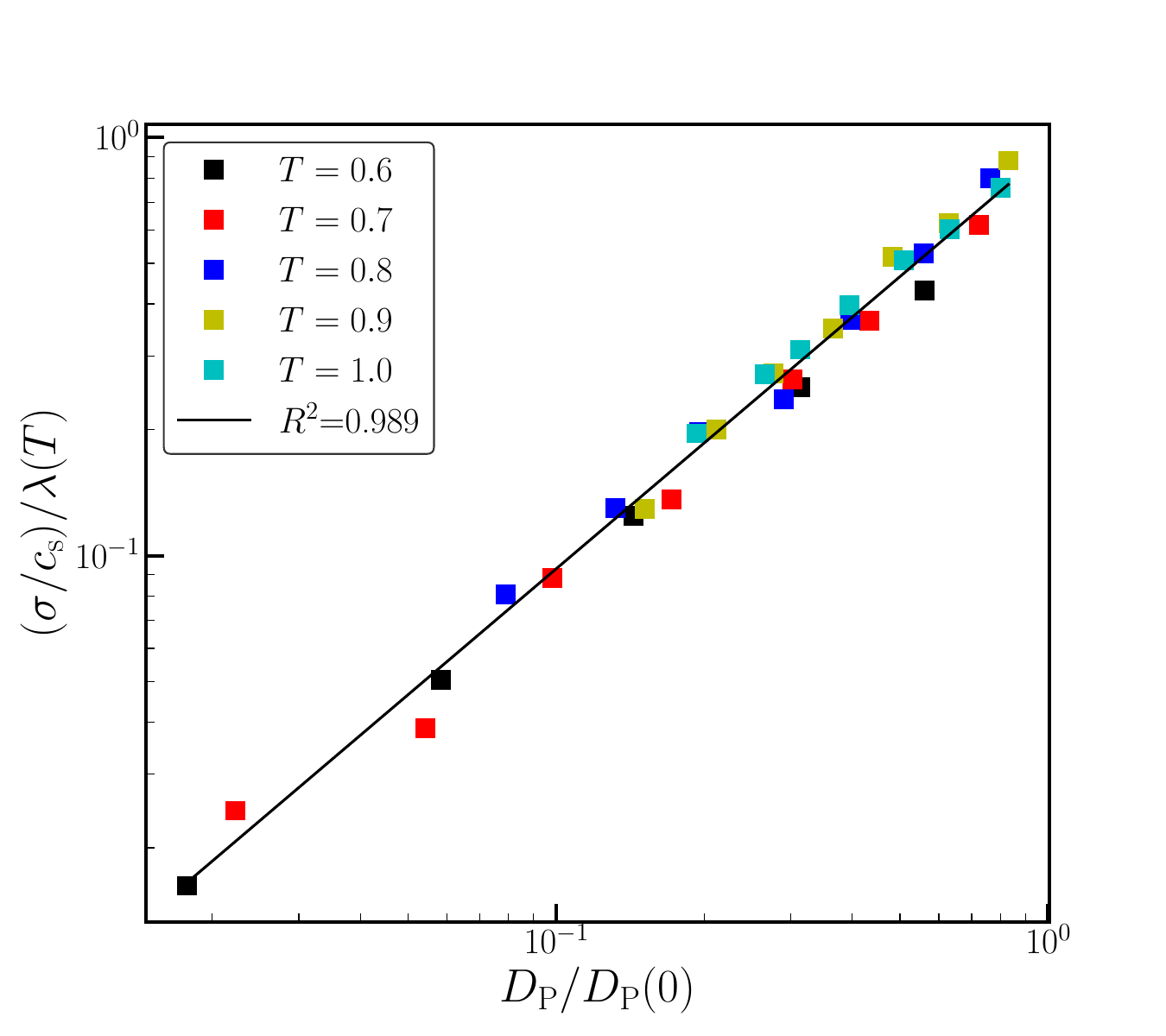}
\caption{ \protect\subref{fig:cond_cs_T} Conductivity, $\sigma$, as a function of salt concentration, $c_{\text s}$, for various temperatures, \protect\subref{fig:sigma_D_ratios} Ratio of the specific conductivities correlates well with the ratio of polymer center-of-mass diffusivity.}
\end{figure}

In \Cref{fig:cond_cs_T}, we present the simulation data (symbols) for the ionic conductivity as a function of salt concentration for various temperatures; in the same figure, we show the results of fitting using \cref{eq:sigma_Dcom} (lines). 
As expected, the ionic conductivity decreases as the temperature is decreased, while the characteristic non-monotonic dependence is found for all temperatures. The maximum in the conductivity shifts to lower concentrations as the temperature decreases, in agreement with experiments. \cite{hoffman_temperature_2021} 
All these trends are captured by \cref{eq:sigma_Dcom}, which yields nearly quantitative agreement with the simulation data. 

As a global measure of the quality of fitting the data using \cref{eq:sigma_Dcom}, we plot the ratio of the specific conductivity $\sigma /c_\text{s}$ to its infinite dilution value $\lambda(T)$ vs. the ratio of the polymer diffusivity $D_\text{{P}}(c_\text{s})/D_\text{{P}}(0)$ on a log--log scale. All the data fall on the same curve, with an R-squared value of 0.989. This result is strong indication that the salt concentration dependence in the ionic conductivity is governed by the segmental dynamics of the polymer for all the temperatures and salt concentrations examined in our work. 

In the literature, the temperature dependence of the ionic conductivity in amorphous polymeric materials is widely fitted to the VFT equation: \cite{thomas_electronic_2021,albinsson_ionic_1992,ratner_conductivity_1989} 

\begin{equation} 
\sigma = A\text{exp} \left( -\dfrac{E_{ a}}{T-T_0} \right)
\label{eq:VFT}
\end{equation}
\noindent where $A$ and $E_a$ are both fitting parameters, and  
the Vogel temperature $T_0$ is typically taken to be 50 K below $T_{\text g}$, or treated as a fitting parameter. 
In Section 6 of the SI, we show the simulation data for the conductivity and the VFT fitting as a function of $1/T$ for different salt concentrations.  While VFT provides a good fit to simulation data, we emphasize that the VFT parameters have to be fitted {\em for each salt concentration}, and their resulting dependence on the salt concentration exhibits behaviors that are difficult to justify on physical grounds, as shown and discussed in SI. 
We believe the unusual behaviors of these fitting parameters with salt concentration, also reported in experimental studies,\cite{chintapalli_structure_2016} raise doubt about the physical basis in the VFT for describing the conductivity data in salt-doped polymers.

In summary, by examining the salt concentration effects on the glass transition of the polymer, we propose a modified VFT equation for the polymer segmental mobility.  Using the M-VFT equation, we provide a simple formula, \cref{eq:sigma_Dcom}, that accurately captures the conductivity data for all the temperatures and salt concentrations studied in our simulation, and affords a straightforward physical interpretation. The complex concentration and temperature dependence in \cref{eq:sigma_Dcom} cannot be captured by either \cref{eq:Sigma_scale} or \cref{eq:VFT} without using parameters that must be fitted for each temperature or concentration and that lack consistent physical interpretation. 
Interestingly, while lowering the temperature and increasing the salt concentration both slow down the segmental dynamics, we note that \cref{eq:sigma_Dcom} does not follow a simple salt--temperature superposition as found in other systems such as ionomers. \cite{zhang_glass_2005} 
A possible reason for this difference may be that the salt concentration in the salt-doped polymer systems has a different role from the temperature in that it alters the activation energy in the VFT equation, in addition to changing $T_{\text{g}}$. 
Our study highlights the nontrivial synergistic effects of salt concentration and temperature on the segmental dynamics of polymers and ion conductivity, and the role of the glass transition temperature,  in salt-doped polymer systems.

\begin{acknowledgement}
This research was supported by funding from Hong Kong
Quantum AI Lab, AIR@InnoHK of the Hong Kong Government. 
The authors thank Prof. Lisa Hall and Dr. Kuan-Hsuan Shen for providing the LAMMPS solvation potential source code.  \end{acknowledgement}

\begin{suppinfo}
Simulation Details; Determination of the Glass Transition; Ion--Polymer Coordination; Rouse-Mode Analysis and Friction Coefficient; Specific Conductivity at Infinite Dilution; and VFT Analysis.
\end{suppinfo}

\bibliography{ref}

\end{document}




\tableofcontents

\section{Simulation Details}








Our simulations include 400 polymer chains, each with 30 neutral monomer beads and are conducted using Lennard--Jones (LJ) units.
We choose the LJ parameter $\epsilon$ to correspond to $k_B T$ at $T=400$K. 
Thus, the reduced temperature $T=1.0$ corresponds to $T=400$K.
This correspondence yields a glass transition temperature that falls within the known values of the glass transition temperature for PEO \cite{lascaud_phase_1994}.
Following previous studies, \cite{shen_ion_2020} the salt concentration mapping from LJ units to real units is $2c_{\text s}^{\text{ LJ}} = c_{\text s}^{\text {real}} \equiv c_{\text s}$, where $c_{\text s}$ is defined as the ratio [Li$^+$]/[EO]. 
To compare with experiments, we report the salt concentration in real units. 
The length scale $\sigma$ is taken as  $0.7$nm.
All particles have the same mass $m=1.0$.

All particles interact via the Lennard--Jones potential:
\begin{equation}\label{eq:LJ}
U_{\text{LJ}}(r_{\text{ij}}) =  
    \begin{cases}
     4\epsilon_{\text{ij}}\left[\left(\frac{\sigma_{\text{ij}}}{r_{\text{ij}}}\right)^{12} - \left(\frac{\sigma_{\text{ij}}}{r_{\text{ij}}}\right)^{6} - \left(\frac{\sigma_{\text{ij}}}{r_c}\right)^{12} + \left(\frac{\sigma_{\text{ij}}}{r_c}\right)^{6}\right],    & r_{\text{ij}}\leq r_{\text{c}}  \\
     0, & r_{\text{ij}} > r_{\text{c}} \\
    \end{cases}
\end{equation}
where $r_\text{ij}$ is the distance between two particles. 
We choose the same LJ interaction energy between all pairs $\epsilon_{\text{ij}}=\epsilon$ and set $\sigma_{\text{ij}}$ to be the mean particle size $\sigma_{\text{ij}}=(\sigma_{\text{i}}+\sigma_{\text{j}})/2$. 
To approximate the size difference between the Li$^{+}$ cation, TFSI$^{-}$ anion, and EO monomers  in the PEO/LiTFSI system, we choose  the polymer bead size to be $1.0 \sigma$, the cation diameter $\sigma_+ =0.4\sigma$, and the anion diameter $\sigma_-=1.6\sigma$.
For the ion--ion, ion--monomer and bonded monomer--monomer interactions, the cutoff distance is $r_\text{c}=2^{1/6}\sigma_{\text{ij}}$, corresponding to the repulsive Weeks--Chandler--Andsersen potential \cite{weeks_role_1971}. 
For non-bonded monomer--monomer interactions we choose $r_\text{c}=2.0 \sigma$, so that the interaction includes the attractive portion of the potential, which is essential to reproduce the glass transition phenomenon. \cite{morita_study_2006}
 
The chain connectivity is modeled using the finitely extensible nonlinear elastic potential (FENE) \cite{grest_efficient_1996,kremer_dynamics_1990} between consecutive beads along the chain backbone: 
\begin{equation}\label{eq:FENE}
U_{\text{FENE}}(r_{\text{ij}}) = -\frac{1}{2} K R_0  \text{ln}  \left(  1 - \dfrac{r_{\text{ij}}^2}{R_{0}^2}  \right)  
\end{equation} 
We choose the standard values $K = 30 \epsilon/\sigma^2$ for the spring constant, and $R_0 = 1.5 \sigma$ for the cutoff radius.

The strong ion--ether oxygen interaction in the lithium salt-doped PEO systems is captured by a solvation potential proposed by Hall and coworkers  \cite{brown_ion_2018, shen_molecular_2021}:

\begin{equation}\label{eq:SOLV}
U_{\text{SOLV}}(r_{\text{ij}}) =  
    \begin{cases}
     -S_{\text{ij}} \left[  \left(  \dfrac{\sigma_{\text{ij}}  }{r_{\text{ij}}}  \right)^{4} - \left(  \dfrac{\sigma_{\text{ij}}  }{r_{\text{c}}}  \right)^{4}       \right],    & r_{\text{ij}}\leq r_{\text{c}}  \\
     0, & r_{\text{ij}} > r_{\text{c}} \\
    \end{cases}
\end{equation} 

\noindent Following these authors, we choose $S_{\text{ij}} = 4.33$ and $r_{\text{c}} = 5.0\sigma$.  

Ion--ion interactions are described by the Coulomb potential
\begin{equation}\label{eq:COULOMB}
U_{\text{COUL}}(r_{\text{ij}}) = \dfrac{q_\text{i}q_\text{j}}{4 \pi \varepsilon_0 \varepsilon_\text{r} r_{\text{ij}}}
\end{equation} 

\noindent where $q_\text{i}$ is the charge of ion i, $\varepsilon_0$ is the vacuum permittivity, and $ \varepsilon_{\text{r}}$ is the dielectric constant of the polymer host medium. For the temperature range studied we choose  constant $\varepsilon_r= 7.5$, corresponding to the dielectric constant of PEO. \cite{shen_ion_2020}

All simulations were performed on the LAMMPS platform.\cite{plimpton_fast_1995} 
We used the velocity-Verlet algorithm with a time step of $0.005\tau$; where $\tau$ is the LJ time defined by $\sigma \sqrt{m/\epsilon}$.
The Coulomb interactions were evaluated using a particle--particle particle--mesh Ewald solver.\cite{brown_implementing_2012}
To control the temperature and pressure of the system we use the Nosé--Hoover barostat and thermostat, \cite{martyna_explicit_nodate} with a damping constant of $1.0\tau$.
The pressure is set to $p=0.0 \varepsilon\sigma^{-3}$. \cite{hsu_coarse-grained_2019}
For all our simulations, we prepared fully equilibrated systems at $T=1.0$ for the range of salt concentrations studied.  To examine the temperature dependence and glass transition, 
we took these equilibrated configurations for the different salt concentrations at $T=1.0$ and applied a constant cooling rate \cite{li_glass_2021}  $\Delta T/\Delta t = -0.5\times  10^{-7} /\tau$ to the desired temperatures.  
Data were collected by further equilibration at each of the desired temperature values. The results were obtained by block averaging with at least four statistically uncorrelated blocks, where the polymer and ions reached the diffusive regime in each block.

\section{Determination of the Glass Transition}

\begin{figure}[H]
\centering
\subfloat[]{
\label{fig:V_T}}
  \includegraphics[width=3.25 in]{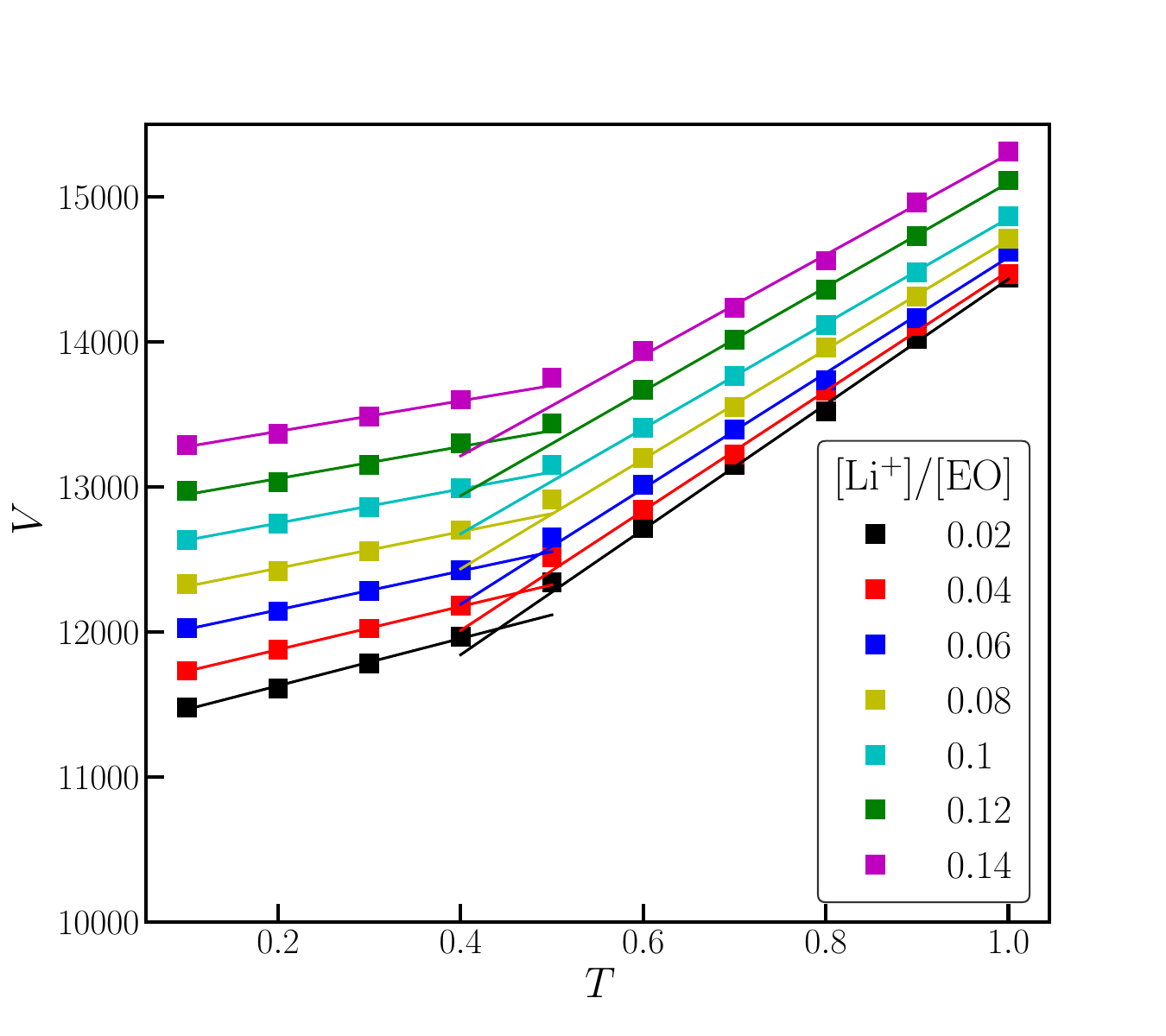}
\subfloat[]{
\label{fig:Vp_cs}}
  \includegraphics[width=3.25 in]{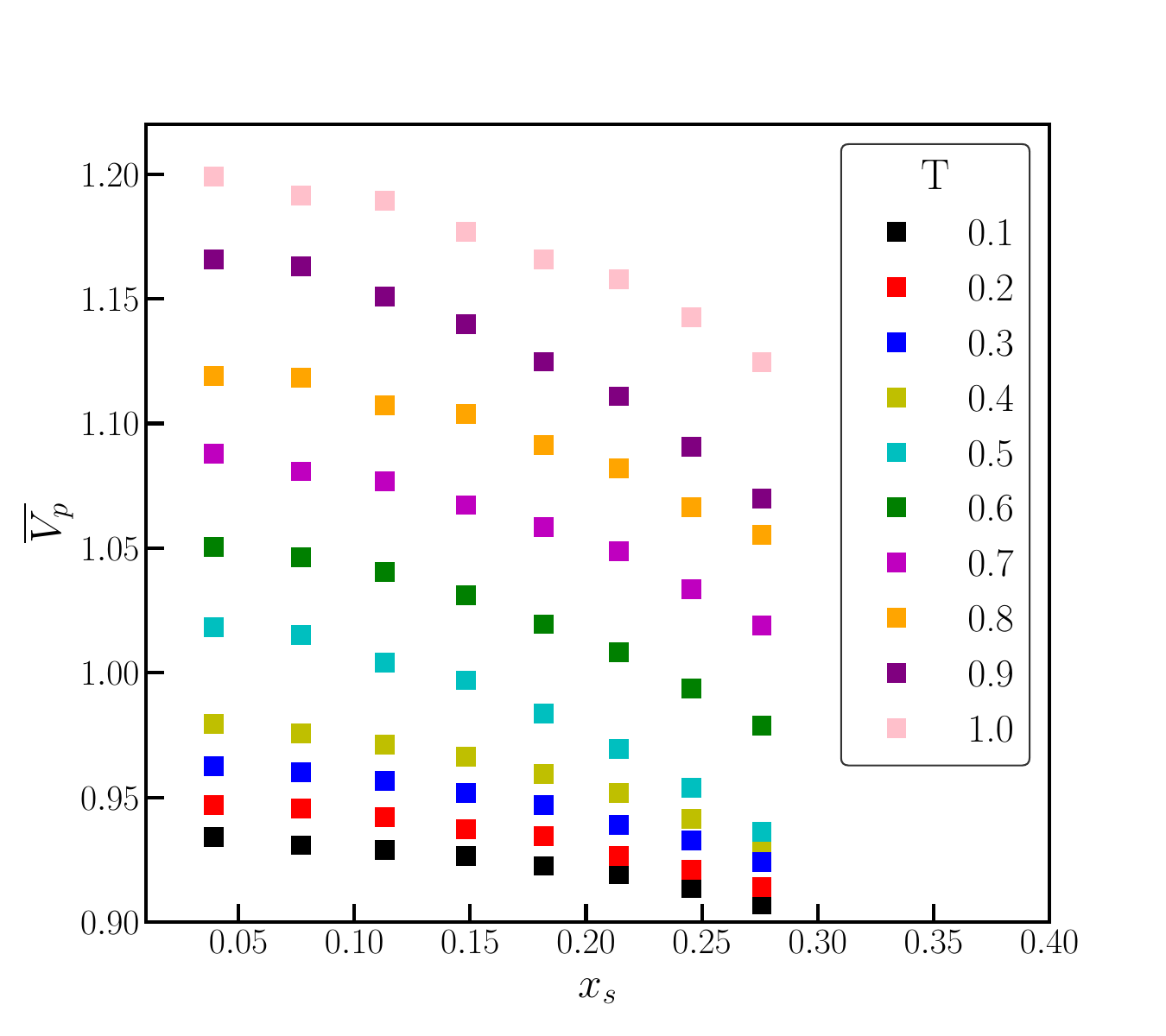}
\caption{\protect\subref{fig:V_T} System Volume, $V$, as a function of temperature, $T$ \protect\subref{fig:Vp_cs} Polymer partial molar volume as a function of salt mole fraction}
\end{figure}

In \Cref{fig:V_T}, we show the system volume, $V$, as a function of temperature, $T$, for various salt concentrations.
Each set of data exhibit two distinct slopes. By linearly fitting the volume data at low and high temperatures, we determine $T_\text{g}$ from the crossing of the two straight lines. 
The increase in $T_\text{g}$ with salt mole fraction can be attributed to the strong ion–polymer complexation, that leads to restricted motion of the polymer backbone, and to the decreased partial molar volume of
the polymer as shown in \Cref{fig:Vp_cs}. 
The monotonic behavior of $\overline{V_p}$ is found for all studied temperatures (below and above $T_\text{g}$).

\section{Ion--Polymer Coordination}

\begin{figure}[H]
\centering
\includegraphics[width=3.25 in]{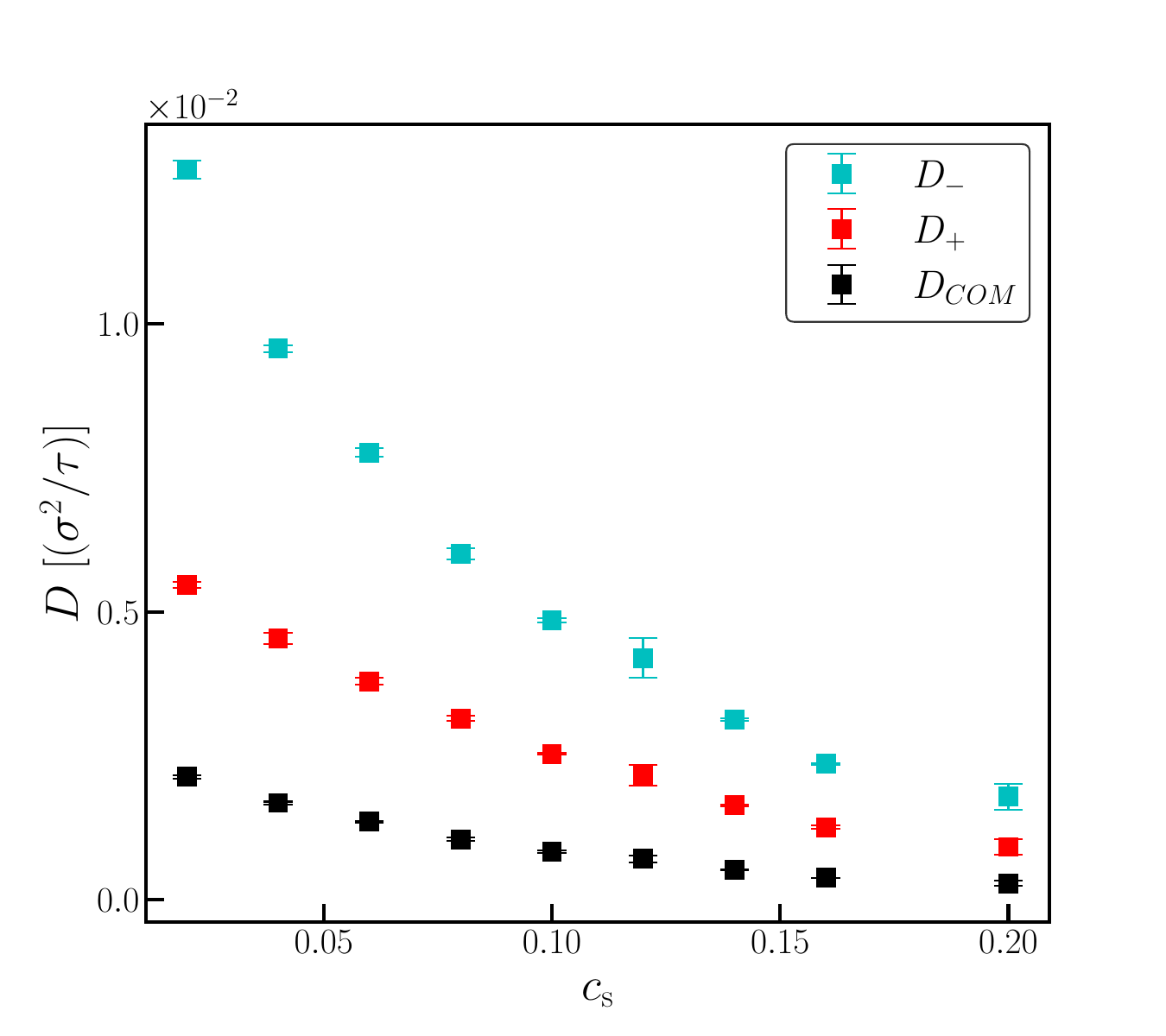}
\caption{Self-diffusion coefficient, $D$, as a function of salt concentration, $c_\text{s}$, for the anion, cation, and polymer center of mass, at temperature $T=1.0$.}
\label{fig:D_cs}
\end{figure}

In \Cref{fig:D_cs}, we show that the self-diffusion coefficients of the cation, anion, and polymer center--of--mass decrease with increasing salt concentration, $c_{\text s}$. 
This is attributed to the slowdown of the segmental dynamics due to the strong ion--polymer coupling. \cite{borodin_mechanism_2006}

\begin{figure}[H]
\centering
\subfloat[]{
\label{fig:gr_lowc}}
  \includegraphics[width=3.25 in]{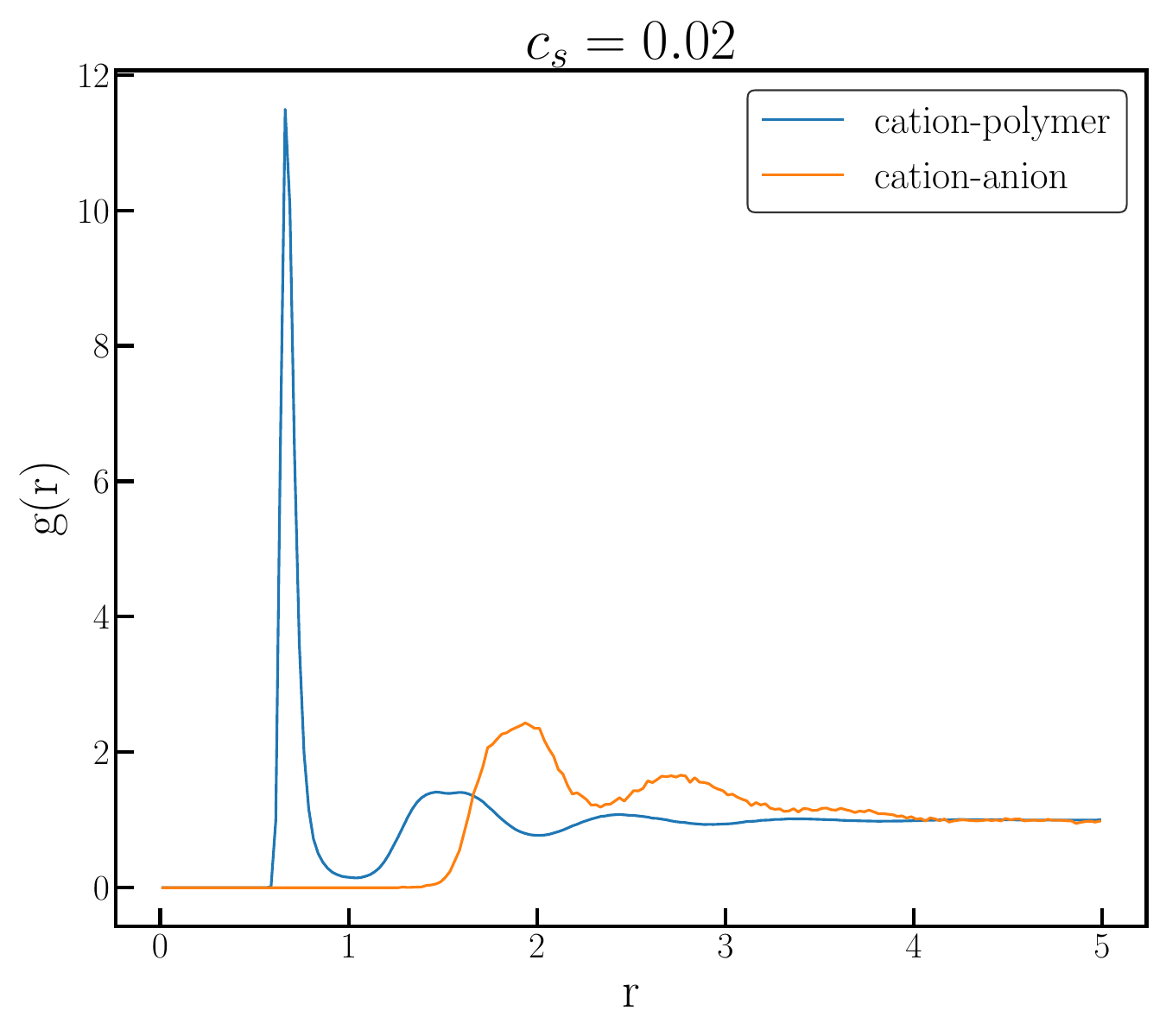}
\subfloat[]{
\label{fig:gr_highc}}
  \includegraphics[width=3.25 in]{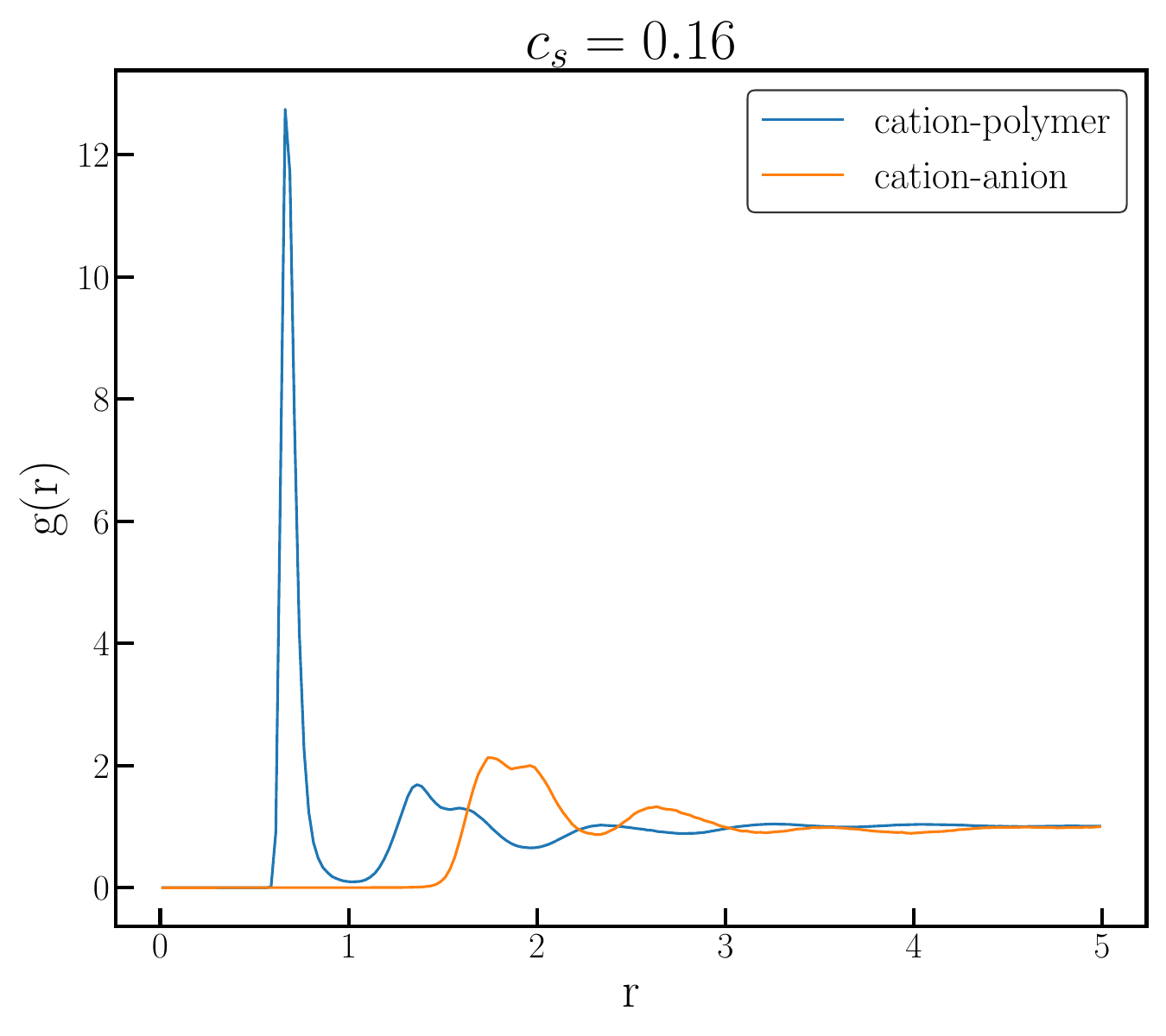}
\caption{ The cation--polymer and cation--anion radial distribution functions at \protect\subref{fig:gr_lowc} $c_{\text s}=0.02$ and  \protect\subref{fig:gr_highc} $c_{\text s}=0.16$}
\label{fig:gr}
\end{figure}

In  \Cref{fig:gr}, we show the cation--polymer and the cation--anion radial distribution function $g(r)$, for $c_{\text s} = 0.02$ and $c_{\text s}=0.16$.
For both concentrations, the peak height of the cation-polymer $g(r)$ is much higher than the peak height of the cation-anion $g(r)$. 
This result supports that the strong cation--polymer interaction leads to seclusion of the cation from forming ion pairs and clusters with the bulky anion in our model system.

\section{Rouse-Mode Analysis and Friction Coefficient}

\begin{figure}[H]
\centering
\subfloat[]{
\label{fig:modes_lowT}}
  \includegraphics[width=3.25 in]{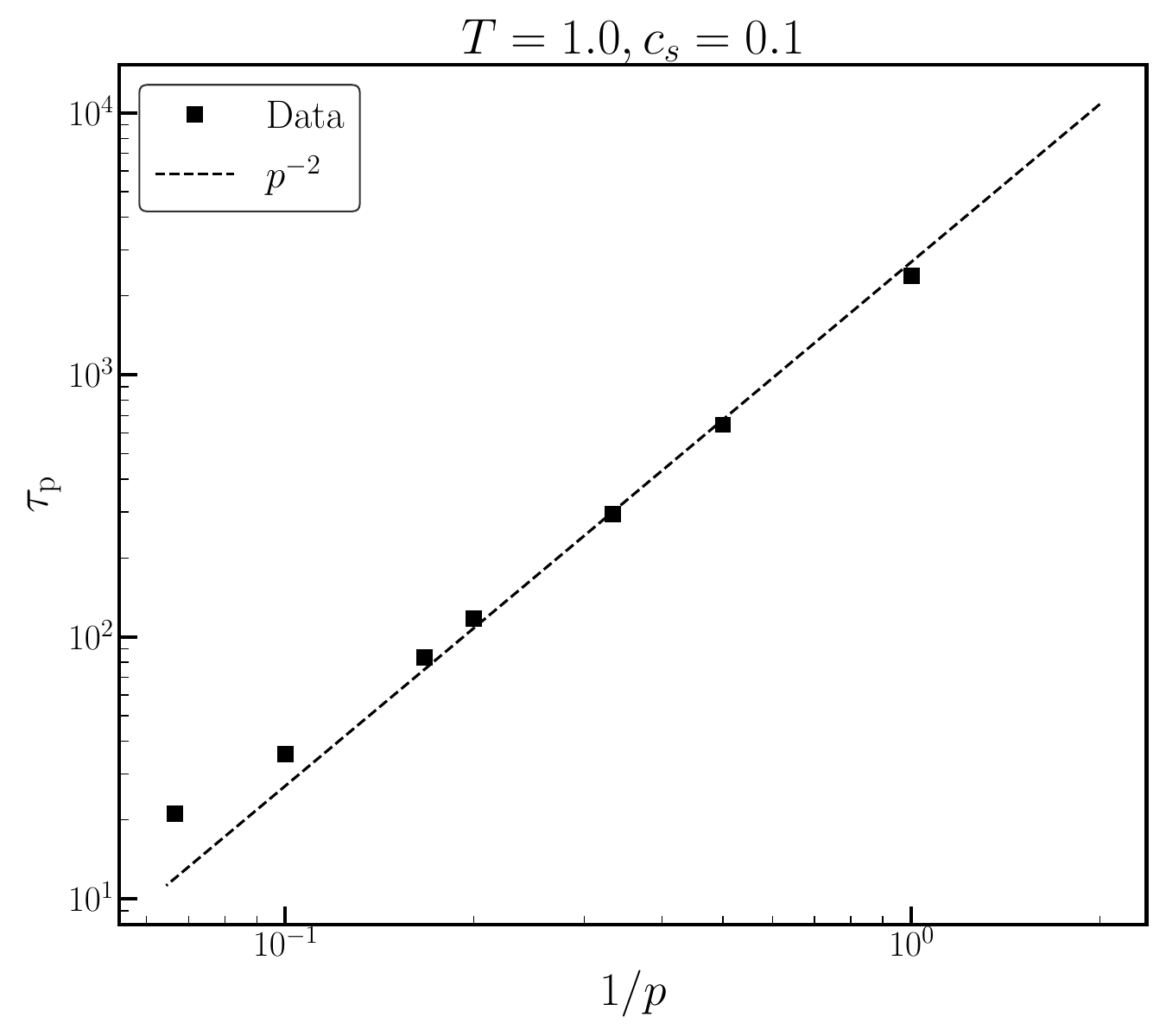}
\subfloat[]{
\label{fig:modes_highT}}
  \includegraphics[width=3.25 in]{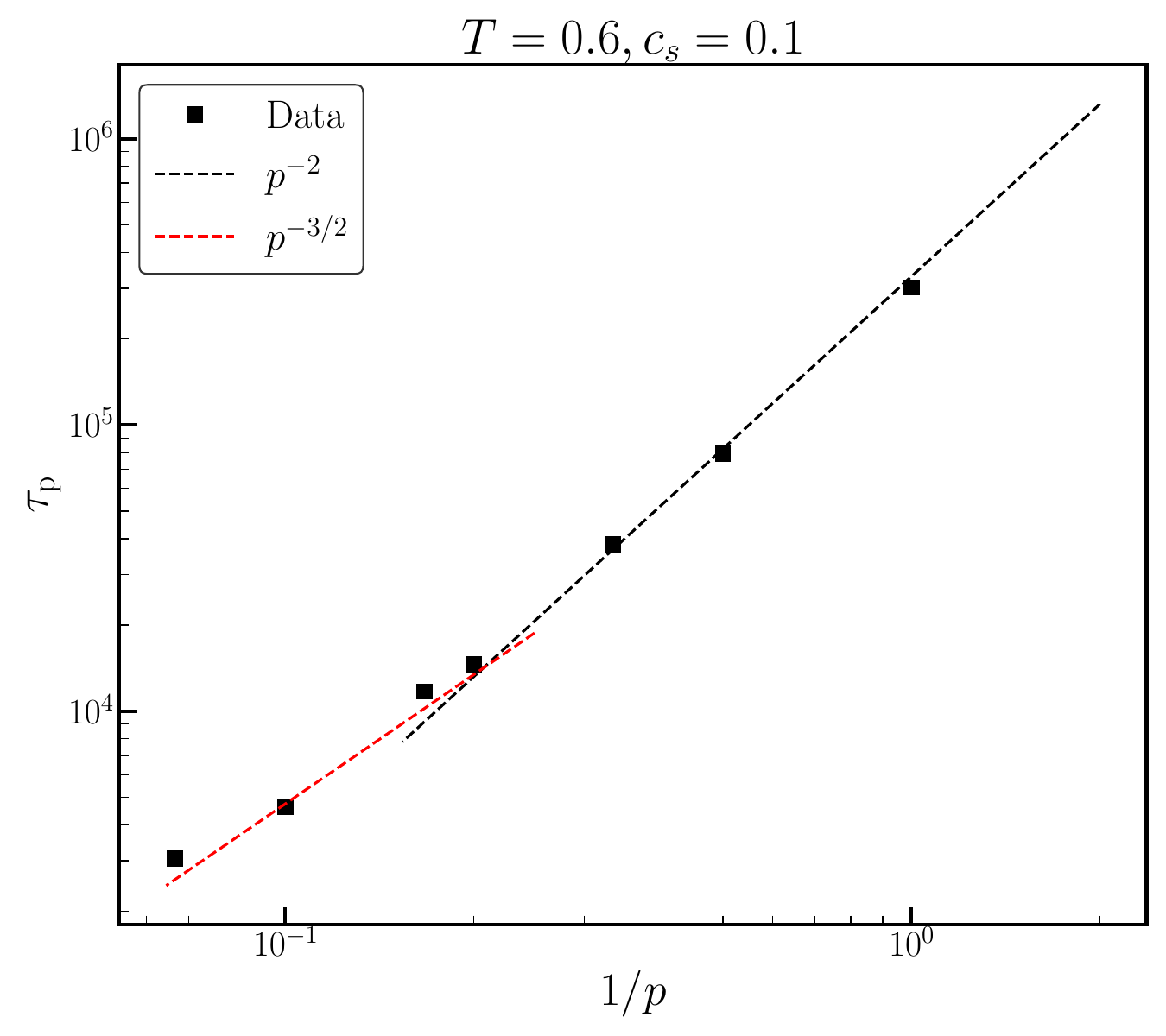}
\caption{  Effective relaxation times plotted as a function of 1/p for \protect\subref{fig:modes_lowT} $T=1.0$ and  \protect\subref{fig:modes_highT} $T=0.6$}
\label{fig:modes}
\end{figure} 

In \Cref{fig:modes} we show the normal modes relaxation times, $\tau_{\text p}$, at $T=1.0$ and $T=0.6$, for salt concentration $c_{\text s} = 0.1$.
The normal mode analysis is described in Refs \citenum{hsu_detailed_2017,kalathi_rouse_2015}.
We find that at $T=1.0$, $\tau_{\text p}$ follows the expected $p^{-2}$ Rouse scaling. 
However, at $T=0.6$ there is a clear deviation from the Rouse prediction. 
Specifically, there is a transition at some intermediate mode from  $p^{-2}$ to  $p^{-3/2}$. The relationship between the location of the crossover and the temperature or salt concentration will be explored in future work.

\begin{figure}[H]
\centering
\includegraphics[width=3.25 in]{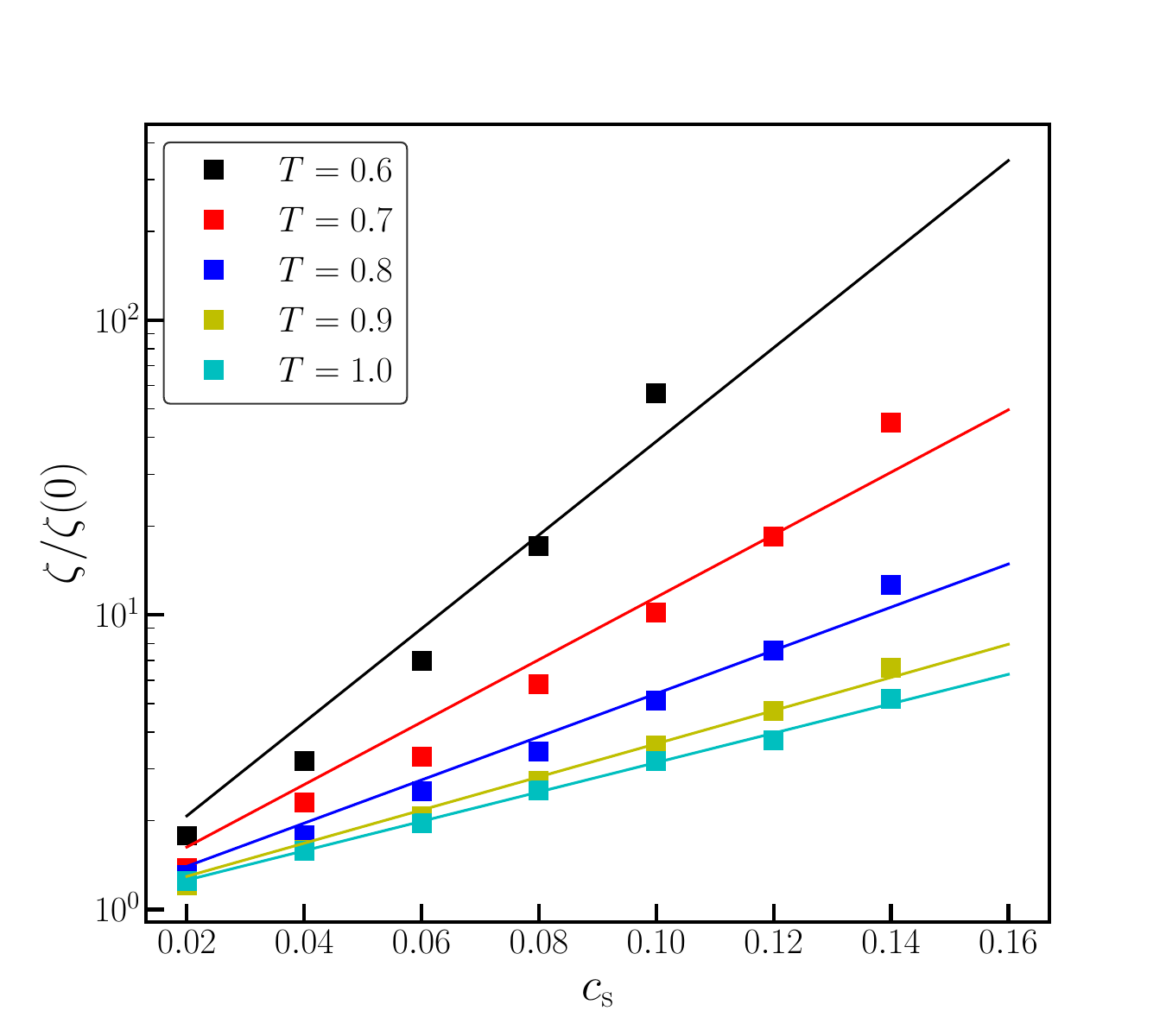}
\caption{Normalized friction coefficient, $\zeta$, as
a function of salt concentration, $c_{\text s}$, for various temperatures.}
\label{fig:zeta_cs}
\end{figure}

In \Cref{fig:zeta_cs}, we show the monomeric friction coefficient calculated from the Rouse model: 
\begin{equation}\label{eq:Rouse}
\zeta =  \dfrac{k_{\text{B}}T}{ND_{\text{P}}} 
\end{equation} 
With decreasing temperature, the salt concentration dependence of $\zeta$ changes from exponential to super-exponential.
Due to this deviation, in the main text we have quantified the segmental dynamics using the modified-VFT equation for the polymer diffusion coefficient, instead of an exponential dependence on salt concentration for $\zeta$.

\section{Specific Conductivity at Infinite Dilution}

\begin{figure}[H]
\centering
\includegraphics[width=3.25 in]{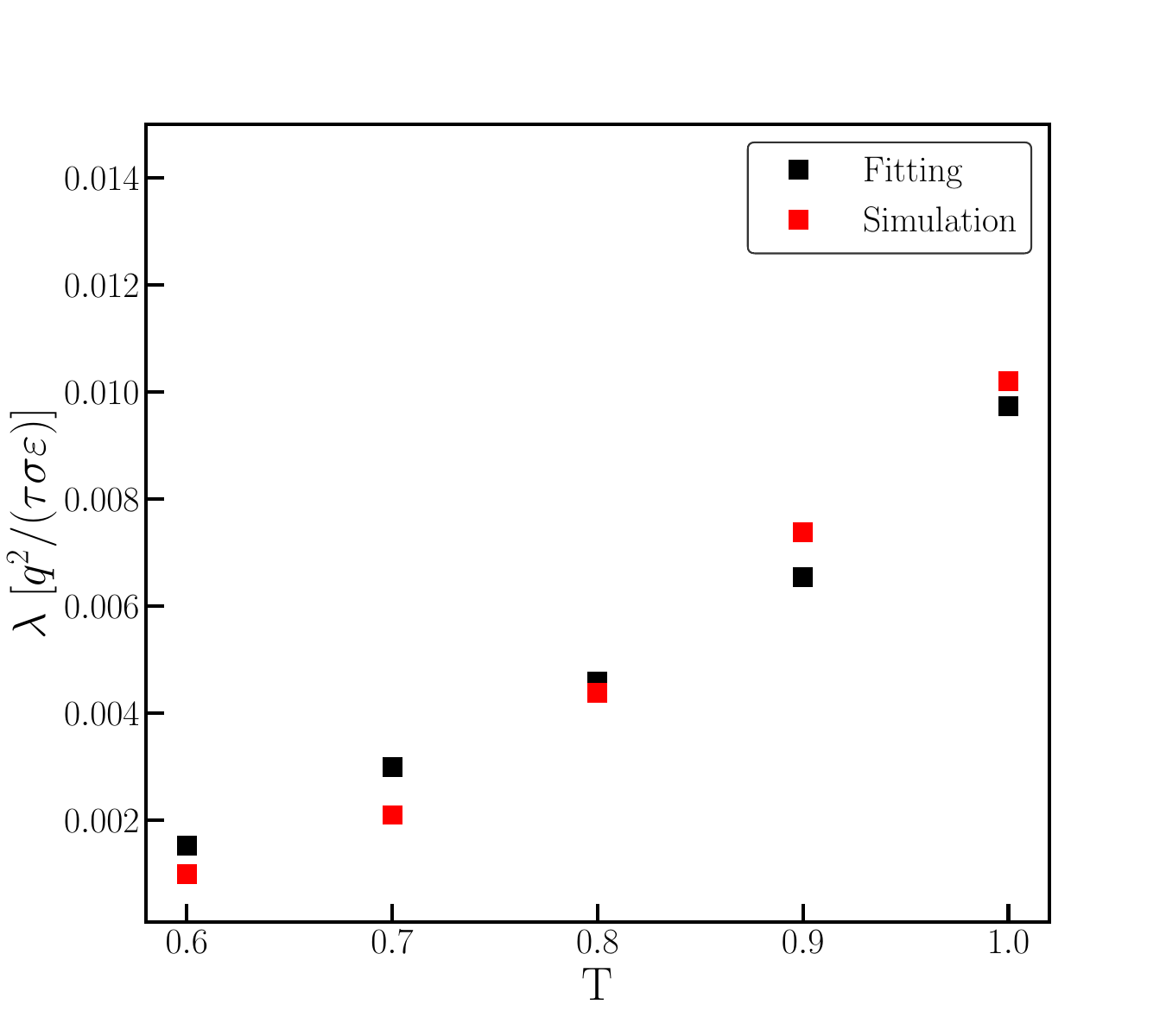}
\caption{Specific Conductivity at infinite dilution as a function of temperature, calculated from fitting eq. 6 and conducting simulations at very low concentration}
\label{fig:specific_sigma}
\end{figure}

In \Cref{fig:specific_sigma}, we show the specific conductivity at infinite dilution, $\lambda$, as a function of temperature. 
We determine $\lambda$ by fitting our conductivity data to eq. 6, and by conducting molecular simulations at very low salt concentrations ($c_{\text s} = 0.00016$). 
We find good agreement between the two methods for all the studied temperatures.

\section{VFT Analysis}

\begin{figure}[H]
\centering
\includegraphics[width=3.25 in]{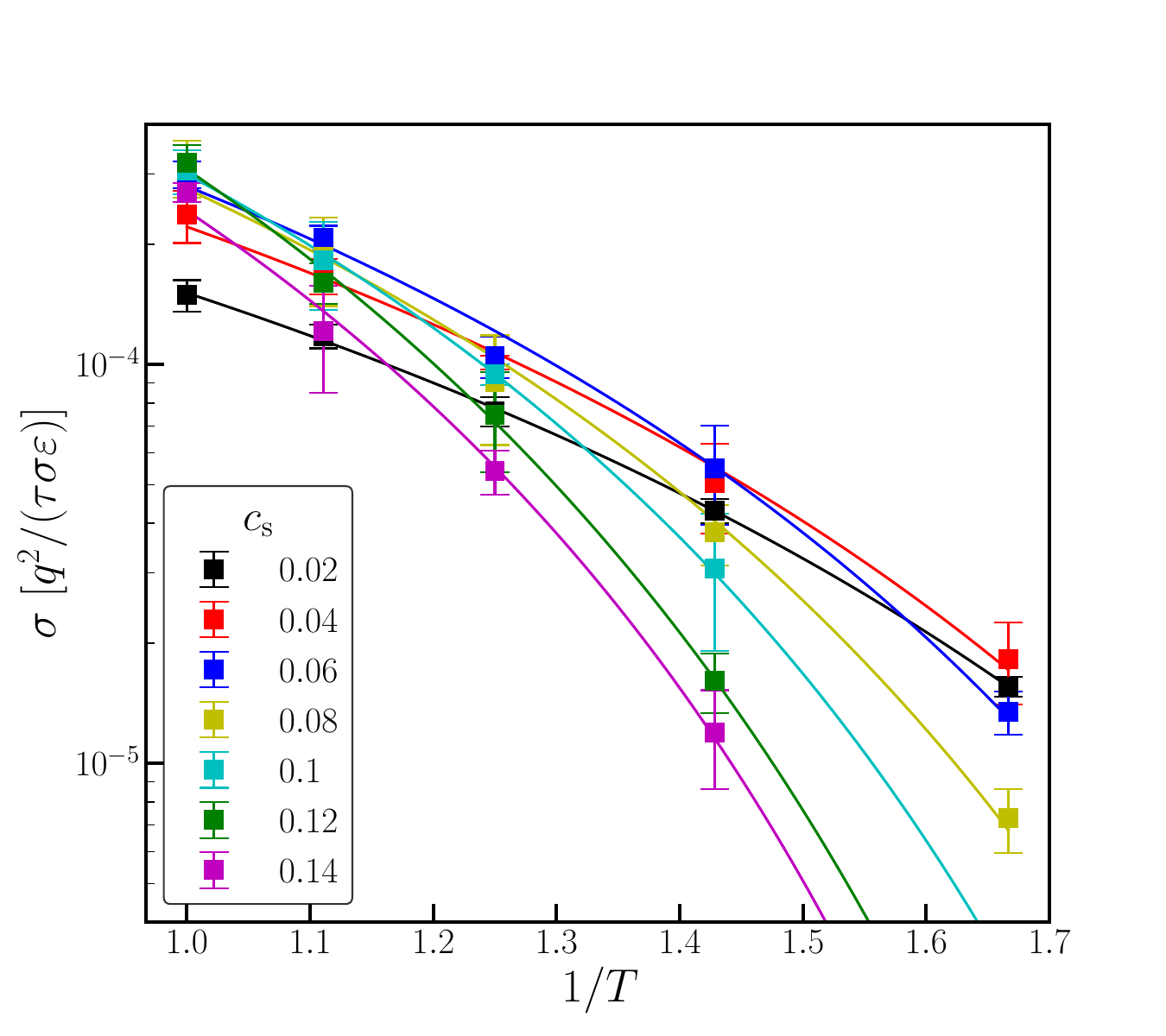}
\caption{Ionic conductivity as a function of 1/T for various salt concentrations. The curves are results from fitting using scheme 1; the results are indistinguishable from those using schemes 2 and 3.}
\label{fig:sigma_T}
\end{figure}

In \Cref{fig:sigma_T}, we present the ionic conductivity as a function of $1/T$ for various salt concentrations.  The symbols are the simulation data, and the lines are the results of fitting using 
the VFT equation:
\begin{equation} 
\sigma = A\text{exp} \left( -\dfrac{E_{ a}}{T-T_0} \right)
\label{eq:VFT}
\end{equation}
\noindent where $A$ is a prefactor commonly associated with the charge carrier concentration, $E_{ a}$ is a pseudoactivation energy related to the segmental relaxation, and $T_0$ is the equilibrium glass transition temperature, typically taken 50 K below $T_\text{g}$. \cite{thomas_electronic_2021,albinsson_ionic_1992,ratner_conductivity_1989} 
We fit our simulation data using 3 different schemes: (1) We consider $T_0 = T_{\text g}-0.125$, where $T_{\text g}$ is the glass transition temperature shown in Fig. 1 of the main text and 0.125 corresponds to 50 K in reduced units; 
(2) we consider $T_0$ to be predetermined from fitting the polymer diffusivity in eq. 6 of the main text; 
(3) $T_0$ is taken as a free fitting parameter.
In all methods, $A$ and $E_{a}$ are considered fitting parameters.

\begin{figure}[H]
\centering
\subfloat[]{
\label{fig:A_cs}}
  \includegraphics[width=3.25 in]{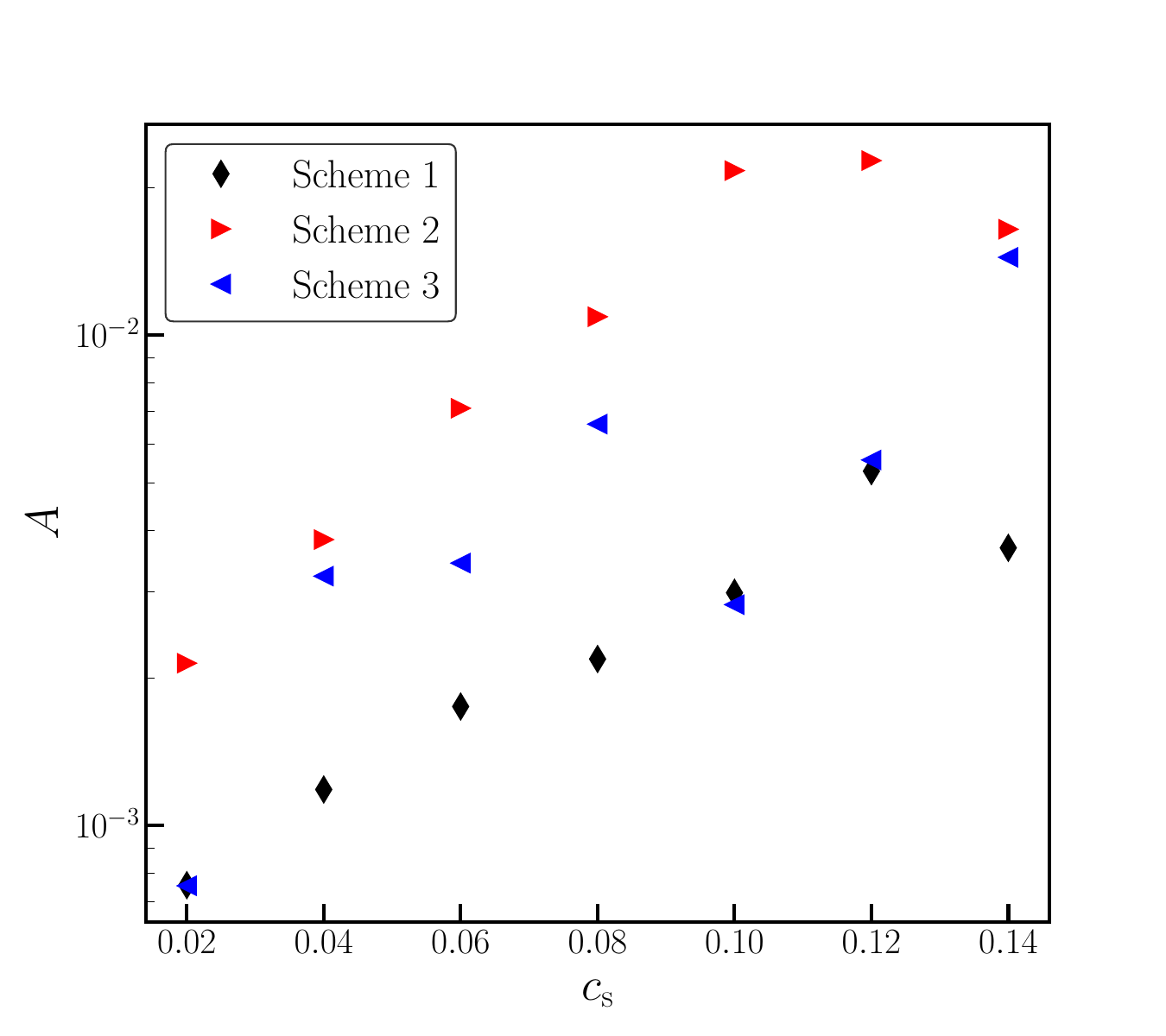}
\subfloat[]{
\label{fig:Ea_cs}}
  \includegraphics[width=3.25 in]{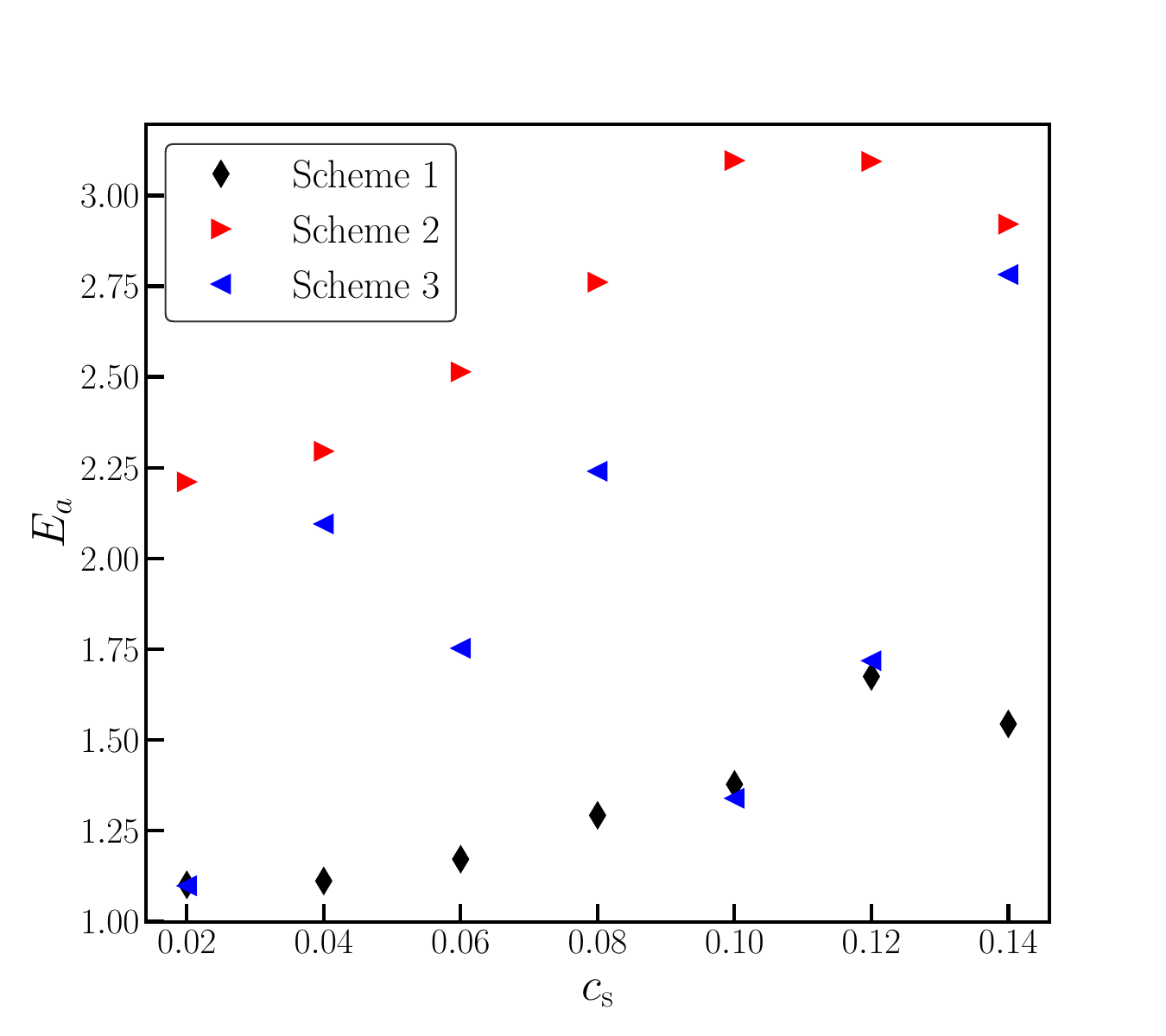}
\subfloat[]{
\label{fig:Tg_cs_SI}}
  \includegraphics[width=3.25 in]{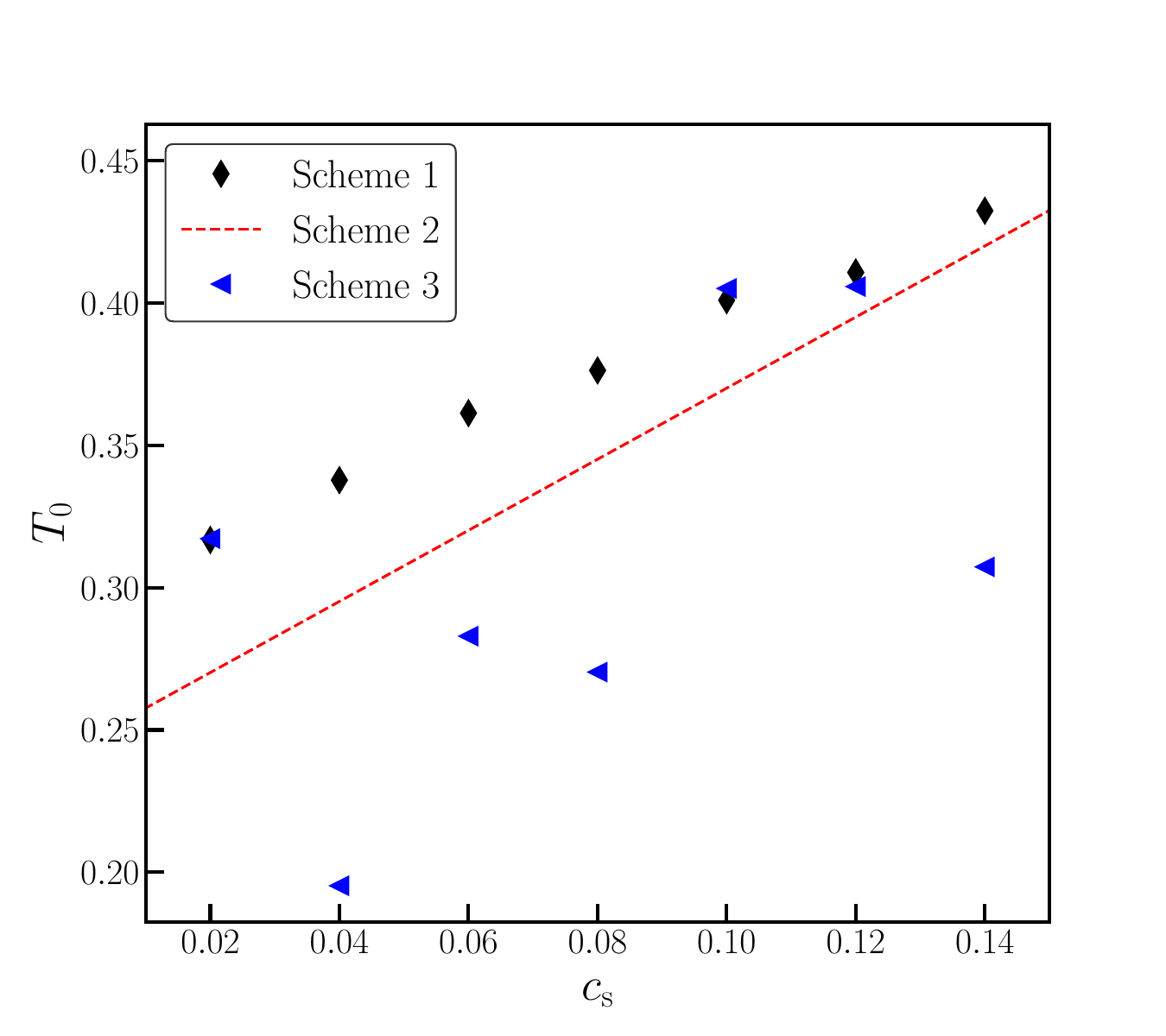}
\caption{ VFT parameters, \protect\subref{fig:A_cs} $A$  \protect\subref{fig:Ea_cs} $E_a$ and  \protect\subref{fig:Tg_cs_SI}   $T_{\text 0}$ calculated as a function of salt concentration, $c_{\text s}$, using the three different fitting methods. 
} 
\label{fig:VFT_params}
\end{figure}

In Figure \ref{fig:VFT_params}, we show the parameters $A$, $E_a$, and $T_0$ as function of salt concentration, $c_{\text{s}}$. 
We observe that for fitting schemes 1 and 2, both $A$ and $E_a$ exhibit a non-linear and non-monotonic dependence on $c_{\text{s}}$.  The non-monotonic behavior in both $A$ and $E_a$ are difficult to justify on physical grounds, given that our model does not have strong ion--ion correlations. 
Fitting scheme 3 yields a rather irregular dependence of $T_0$ on $c_{\text{s}}$, as shown in Figure \ref{fig:VFT_params}(c). 
The fitted values of $T_0$ fails to have any meaningful connection to the glass transition temperature, and likewise we find unphysical fluctuations in $A$ and $E_a$. The fact that three different fitting schemes with widely different values for the fitting parameters can yield nearly identical results to fit the conductivity data is strong evidence that the VFT fitting for the conductivity in salt-doped polymers lacks solid physical basis.
\bibliography{ref}